\documentclass{jfm}
\usepackage[mylang]{appendix}
\usepackage{jfmcaps,epsfig,psfig}
\usepackage[english]{babel}

\ifCUPmtlplainloaded
\else
\fi
\ifCUPmtlplainloaded

\fi
\ifCUPmtlplainloaded
  \font\bit = mtmib10 at 10.5pt \skewchar\bit ='177  % bold math italic
\else
  \font\bit = cmmib10 \skewchar\bit ='177  % bold math italic
\fi
\ifCUPmtlplainloaded
\else
  \font\tenbmi=cmmib10 at 10pt  \skewchar\tenbmi ='177
  \font\sevenbmi=cmmib10 at 7pt \skewchar\sevenbmi ='177
  \font\fivebmi=cmmib10 at 5pt  \skewchar\fivebmi ='177

  \newfam\bmifam
  \textfont\bmifam=\tenbmi
  \scriptfont\bmifam=\sevenbmi
  \scriptscriptfont\bmifam=\fivebmi
  
\fi
\newsavebox{\thalfbox}
\sbox{\thalfbox}{$\textstyle\frac{1}{2}$}

\newsavebox{\shalfbox}
\sbox{\shalfbox}{$\scriptstyle\frac{1}{2}$}

\newsavebox{\squartbox}
\sbox{\squartbox}{$\frac{1}{4}$} %RM removed scriptstyle

\newsavebox{\etbox}
\sbox{\etbox}{\boldmath$\eta$}
\newsavebox{\astrutbox}
\sbox{\astrutbox}{\rule[-5pt]{0pt}{20pt}}

\mathchardef\varLambda="0103

\ifCUPmtlplainloaded
\else
\fi
\def\etal{\mbox{\it et al.\ }}

\title[On The Miscible Rayleigh-Taylor instability]
{On the miscible Rayleigh-Taylor instability: two and three dimensions}

\author[Young, Tufo, Dubey, and Rosner]
{Y.-N.\ns Y\ls O\ls U\ls N\ls G$^1$,
H. T\ls U\ls F\ls O$^2$,
A. D\ls U\ls B\ls E\ls Y$^1$,
and \ns
R. R\ls O\ls S\ls N\ls E\ls R$^1$}

\affiliation{$^1$Department of Astronomy and Astrophysics, University of
Chicago,
Chicago, IL 60637, USA\\
$^2$Department of Computer Science, University of
Chicago, IL 60637, USA\\[\affilskip]}

\pubyear{2001}
\journal{J. Fluid Mech.}
\volume{477}
\pagerange{377--408}
\date{15 December 1999 and in revised form 5 December 2000}
\shorttitle{On the miscible Rayleigh-Taylor instability}

\begin{document}

\maketitle

\begin{abstract}
We investigate the miscible Rayleigh-Taylor (RT) instability in both 2
and 3 dimensions using direct numerical simulations,
where the working fluid is assumed incompressible under the 
Boussinesq approximation. 
We first consider the case of randomly
perturbed interfaces. With a variety of diagnostics, we develop a
physical picture for the detailed temporal development of the mixed
layer: We identify three distinct evolutionary phases in the development of
the mixed layer, which can be related to detailed variations in the
growth of the mixing zone. Our analysis provides an explanation
for the observed differences between two and three-dimensional
RT instability; the analysis also leads us 
to concentrate on the RT models  
which (1) work equally well for both laminar and turbulent flows,
and (2) do not depend on turbulent scaling 
within the mixing layer between fluids.  
These candidate RT models
are based on point sources within bubbles (or plumes) and
interaction with each other (or the background flow).
With this motivation, we examine the evolution of 
single plumes, and relate our numerical results (of single plumes)
to a simple analytical
model for plume evolution. 

\end{abstract}

\section{Introduction}

The phenomenology of mixing due to the Rayleigh-Taylor instability, the
instability of an interface separating fluids of different densities
subject to gravity (\cite{Chandra} and references therein), can be
summarized as follows: Bubbles (spikes) of lighter (heavier) fluid
penetrate into the heavier (lighter) fluid, and leave behind them a
region of mixed fluid. As the instability enters the nonlinear regime,
fluid motions in this mixing region become highly irregular. The
envelope, or edge, of the mixing region is observed to be defined by fast
large-scale motion, which tend to be dominated by the merging of
expanding smaller bubbles (spikes). The irregular fluid motions within
the mixing zone are often regarded as ``turbulent" in the sense that the
flow is chaotic in the wake of the bubbles (spikes).

A principal focus of experimental and theoretical study has been the
general properties of this mixing zone, whose broadening in time is
commonly characterized by the ``envelope velocity" $(V_e)$ and the
penetration lengths $h_b$ and $h_s$ (for bubbles and spikes,
respectively). Using simple analytical models for the interpenetration of
the two fluids, one can show (\cite{Youngs84} and references therein)
that in the nonlinear regime $V_e$ is proportional to time $(t)$, and that
the penetration depths $h_{b,s}$ are proportional to $t^2$ and depend
linearly on the gravitational acceleration $g$ and the Atwood number $A$
[$\equiv (\rho_1-\rho_2) / (\rho_1+\rho_2)$, where $\rho_{1,2}$ is the
density of the heavier (lighter) fluid, respectively], i.e.,

\begin{equation}
\label{scaling_law}
h_{b,s} = \alpha_{b,s} gAt^2\;.
\end{equation}
$\alpha$ can be thought of as a measure of the efficiency of potential
energy release; experimental measurements of $\alpha$ give results in the
range of $0.03\sim 0.06$ (see \cite{Sharp84, Read84} and \cite{Youngs84}).
Most recently, \cite{Schneider_98} show that $\alpha$ lies in the range of
$0.05\sim 0.06$ for an Atwood number of $0.34$; as the Reynolds number
is high ($\sim 10^5$) in these Linear Electric Motor (LEM) experiments,
the instability enters the nonlinear regime within several e-folding times
($\sim 0.9$ ms), so that the above
scaling law seems to hold from the time that the first few measurements
of mixing zone width are made.

There are many possible reasons for the observed variation of $\alpha$
found in the literature, when scaling of the form Eq. (\ref{scaling_law})
is observed during the experiment or simulation. Setting aside problems
such as incompletely controlled experiments or insufficiently resolved
computations, it is important to establish whether the nonlinear
(long-time) evolution of the instability is sensitive to details in the
experiment such initial conditions; and to what extent one can really
talk about universal scaling of the mixed layer width. These questions
have been addressed to some extent in the literature. For example,
\cite{Youngs91} has studied the variation of $\alpha$ with time during
the course of R-T evolution; he reported that larger values of
$\alpha\sim 0.06$ occur during the early phase of the 3-D problem, before
any development of small-scale turbulence; however, as small-scale
motions develop, \cite{Youngs91} reports that the 3-D growth rate slows to
that characteristic of the 2-D case ($\alpha\sim 0.04$). It is also found
that $\alpha$ may depend on the Atwood number
$A$ as $A\rightarrow 1$ (G. Dimonte, private communication). In a
variation on this theme, it has been observed in simulations that
$\alpha$ can depend on dimensionality: $\alpha$ for
three-dimensional calculations is found to be larger
than from 2-D simulations. \cite{Sakagami_Nishihara_90} have
found $\alpha$ for three-dimensional calculations to be at least $4$
times higher than for 2-D simulations (in spherical systems); and
more detailed phenomena can be found in 
\cite{Yabe_Hoshino_Tsuchiya_91}, \cite{Li_93} 
and \cite{He_Zhang_Chen_Doolen_99}.  
\cite{Town_Bell_91} found larger values for $\alpha$ in 3-D only during
the early nonlinear stage, and showed that as small scale 3-D turbulence
developed, the 3-D growth slows to that of the 2-D case, similar to
results reported by \cite{Youngs91}. Other 3-D numerical simulations,
including investigations of single-mode initial perturbation
(\cite{Dahlburg_Gardner_90,Tryggvason_Unverdi_90,JKane98}) and more
general initial perturbations (\cite{Cook98}), obtain a range of
values for $\alpha$ generally restricted to $0.05\sim 0.06$. 
In this paper we also explore the dependence of the RT instability
on dimensionality.  We focus on quantitative comparisons between
two and three dimensions, and these comparisons provide 
more complete insights to the understanding of RT instability.
In the case of experiments conducted with gases, it has also been
shown that the interface perturbation growth can be measurably affected
by mass diffusion effects between the two layers 
\cite{Duff_Harlow_Hirt_62}.
These
experimental and computational results argue strongly that the precise
value of $\alpha$ is sensitive to a variety of details in the experiments,
ranging from the specific nature of the initial state to the
dimensionality of the perturbation. It is the aim of this paper to
initiate a study of this issue: In this paper we consider some of the
effects which influence the actual value of the scaling coefficient
$\alpha$ as the Rayleigh-Taylor instability develops in time; a central
question we address is whether a single scaling regime is always to be
expected during the course of non-linear Rayleigh-Taylor development.

In order to make our study more tractable, we have focused only on the
low Atwood number regime. The advantage of this regime is that numerical
techniques developed for incompressible flows can be readily applied to
flows satisfying the Boussinesq equations; such computations are
substantially more economical than calculations of the fully compressible
problem, and thus allow a more extensive exploration of the parameter
space other than Atwood number. Physically, our calculations correspond
to, for example, cases in which the density contrast is introduced by the
distribution of a scalar field (e.g., salt or temperature), for which the
Atwood number is usually small because the density varies weakly with the
scalar ``concentration". Species or temperature diffusion occurs in such
systems, and is therefore included in our simulations; we are therefore
considering the miscible version of the Rayleigh-Taylor instability; in
this Boussinesq approximation, the Atwood number $A\rightarrow 0$ and yet
$gA$ is a finite constant. An essential element in computations in this
regime is that the calculations remain fully resolved throughout the
evolution of the instability. This important constraint limits the range
of allowable Prandtl number $Pr$ (ratio of viscosity to thermal
diffusivity) or Schmidt number $Sc$ (ratio of viscosity to species
diffusivity) in our calculations; we typically take these ratios to be of
order unity.
\cite{Tryggvason88} and \cite{Aref_Tryggvason_89} have adopted similar
ideas in their earlier studies of immiscible Rayleigh-Taylor instability;
however, in their case, the density is not coupled to a scalar field, and
the buoyancy term in the Boussinesq Navier-Stokes equations (induced by
the sharp density gradient across the interface) is the only term related
to the weak stratification.

There are unfortunately few experimental studies of this regime:
\cite{Linden94} have studied the miscible Rayleigh-Taylor instability
experimentally, and via simulations.
In their experiments, they place a layer of brine above a layer of fresh
water, separated by an aluminum barrier. The experiment is initiated by
sliding the barrier horizontally through one sidewall; the Atwood numbers
in their experiments are in the range of $10^{-3} \sim 10^{-2}$
(as the density contrast due to the brine concentration is small when
compared to the ambient density) and the Schmidt number $Sc$ is $\sim
10^3$. Their experimental estimates for $\alpha$ are around $0.044$.
\cite{Linden94} also performed numerical simulations for an Atwood number
$A=0.2$, and (numerical) Schmidt number of order unity. After imposing
various perturbations at the interface, and determining $\alpha$ as a
function of the ratio of the initial displacement of the fluid to the
perturbation wavelength, they concluded that the smaller this ratio is,
the higher the value of $\alpha$. These authors argued that the removal of
the barrier in the experiment corresponds to a long wavelength
perturbation at the interface, and therefore the comparison between
numerical simulation and experiments makes sense only if the long
wavelength perturbation is included as part of the initial perturbation
in the simulation. By implication, they thus suggest that the value of
$\alpha$ obtained even in the long-time limit can depend on the initial
conditions; in other words, these systems have long-term memory
that is not erased by turbulent motions within the mixing interface.
The dependence of RT instability on the initial conditions has
recently been studied by \cite{Cook2000}.
In this paper, 
we focus on a study of the physical effects which determin $\alpha$,
and in particular investigate effects which lead 
to a departure from simple $t^2$ scaling of the mixing zone depth.  A detailed
study of the scaling law itself is now in progress, and led by G. Dimonte
(the $\alpha$ group).

Our paper is organized as follows: We first formulate the problem, outline
the numerics, and
present evidence for the convergence and accuracy of our calculations. In
the following section (\S\,3) we summarize our results for multi-mode
interface perturbations. We next discuss existing (analytical) models
relevant to our studies (\S\,4), which motivate a synthesis of a new
model; the following section (\S\,5) we explore these ideas in more
detail via numerical studies of isolated plume evolution in both 2-D and
3-D. Our results are discussed in~\S\,6, while our conclusions are
presented in~\S\,7.

\section{Formulation and Methods}

\subsection{Formulation of the problem}
We consider two vertically stacked fluid layers
of different density, governed by the
Boussinesq equations in a rectangular box
(\cite{Chandra}). We
adopt periodic boundary conditions along the horizontal directions;
the top and bottom boundaries are no-flux and no-slip. Under the
Boussinesq approximation, density variations only appear in the buoyancy
term, and are small when compared to the mean density. 
There are a variety of ways of achieving a small density jump across
the horizontal inerface: one may choose gasses of slightly different
mean molecular weight, or a single fluid into which a `contaminant' is
added (such as temperature, sugar or salt in water) which changes
the fluid's density slightly. such vertical density
variation depends linearly on temperature; colder (heavier) fluid sinks
and warmer (lighter) fluid floats as gravity points downwards. Thus, in
our formulation, the density jump across the interface is replaced by 
(for example) a
temperature difference: colder fluid on top of warmer fluid. The
equations for the miscible, incompressible, Rayleigh-Taylor instability
in the Boussinesq limit are then essentially the incompressible
Navier-Stokes equations coupled to an advection-diffusion equation for
the `contaminant', here temperature:
\begin{eqnarray}
\label{3D-NS}
\partial_t\vec{u} + (\vec{u}\cdot\vec{\nabla})\vec{u} &=& -\vec{\nabla}P +
 \nu \nabla^2\vec{u} + \beta g\,  T\hat{g},\\
\label{Divergence}
\nabla\cdot\vec{u} &=&0,\\
\label{ThermalAD_eq}
\partial_t T + (\vec{u}\cdot\vec{\nabla}) T &=& \kappa \nabla^2 T,
\end{eqnarray}
($\nu$ is the kinematic viscosity, $\beta$ is the volumetric expansion
coefficient, and $\kappa$ is the thermal diffusivity.)\ \ We remark that
the initial, unperturbed temperature profile diffuses on the diffusion
time scale. We further remark that one could alternatively treat the case
of weak stratification via a vertically-imposed salinity (or any
other type of concentration) gradient; in this way, one is able to explore
a wide range of Prandtl/Schmidt numbers. There can be a variety of top and
bottom boundary conditions for the scalars, such as fix values or fix
fluxes at the boundaries.  In our simulations, we adopt 
no-flux boundary conditions, which are typical in a run-down experiment.
It is important to note here that since our equations allow for
temperature diffusion, they are equivalent to systems in which the
`contaminant' is salt and allow for saline diffusion, or to systems 
describing two miscible fluids of slightly different density and allow
for mass diffusion.

In our simulations, we are interested only 
in regimes where the thermal diffusion
time is much longer than the dynamical time scales of interest. The
reason for this constraint is as follows: Unlike the case of immiscible
Rayleigh-Taylor instability, the presence of thermal diffusion 
formally does not
allow for a static background temperature profile prior to any
perturbation. Nevertheless, as we perturb the temperature at the
interface, we find that if the diffusivity is sufficiently small, then
the initial growth of the perturbation is exponential and thus mimics the
linear regime in standard stability analyses for which an initial static
equilibrium exists. Thus, long thermal diffusion times are essential if the
dynamics is to be dominated by the Rayleigh-Taylor instability. The same
kind of argument applies in the saline case, or in the case where
mass diffusion can occur.  Thus, we do not expect to see diffusion
effects as reported by \cite{Duff_Harlow_Hirt_62}; and our results
indeed show that at least in the linear regime, our growth rates are the
same as in the inviscid nondiffusive case.

Our problem is defined by a number of 
distinct spatial and temporal scales: the
perturbation wavelength(s) ${\bf \lambda}$ (for single-mode perturbation) or
spatial perturbation spectrum (for multi-mode perturbation), 
the amplitude of the initial
displacement of the fluid at the interface ${\bf a}_{\lambda}$, 
and the initial
interface thickness ${\bf d}$. 
Two timescales are of most interests here:
the dynamical time defined by the free-fall time and the 
scalar diffusion time.
In the present study, we focus on cases
where the dynamical time scale is much smaller than the
diffusion time scale. In this limit, the diffusive interface
between the hot and cold material remains more or less fixed at width
$\bf d$ throughout the evolution; the computation's spatial resolution
is then required to at least resolve this interface throughout the
calculation. Furthermore, we impose perturbations of amplitude
${\bf a}$ such that ${\bf a} \ge {\bf d}$ 
in order to capture the initial perturbation
properly. Finally, the range of Atwood numbers we explore is consistent
with values obtained in laboratory experiments for the miscible RT
instability; and the ratio of viscosity to diffusivity is kept to
values greater than or equal to $1$.

\subsection{Numerics: methods and validation}
We have used two distinct numerical codes to solve the above equations,
based on our desire to 
insure that our results are independent of the computational method
used to solve equations (\ref{3D-NS})-(\ref{ThermalAD_eq}).
We have compared both the mixing zone extent and the evolution
of fluid structures (obtained from these two codes), as to
determine the fidelity of our calculations:
The two codes lead to results within expected range of errors
for the same initial conditions
and boundary conditions.  

\begin{figure}[ht]
\begin{center}
\epsfysize=2.0in
\epsfbox{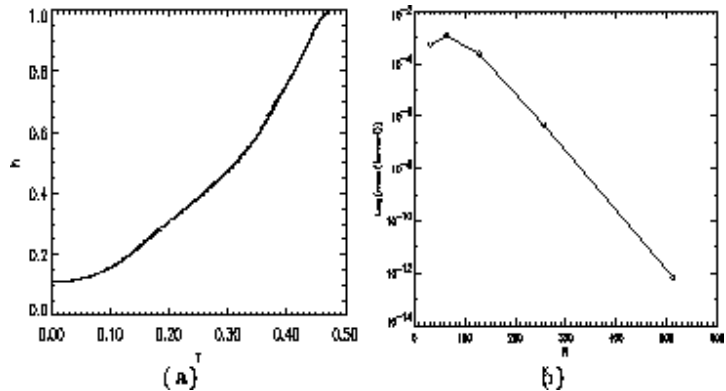}
\end{center}
% \begin{center}
%   \begin{tabular}{c c}
%      {\psfig{file=Code_Comparison_RT.eps,height=2.6in,width=2.5in}} &
%      {\psfig{file=Spatial_res.eps,height=2.5in,width=2.5in}} \\
%      (a) & (b)
%   \end{tabular}
% \end{center}
\caption{Panel (a): Comparison between the spectral-element code (solid line)
         and the spectral code (dashed line) for simulation of single
	 mode perturbation at the interface.  Plotted is the
         mixing zone width $h$ as a function of time $\tau$.
         Panel (b): The spectral code shows exponential convergence for
         a fixed interface thickness of $0.15$ in the single
         mode simulation. The error
         is defined as the deviation from
         the simulation with a resolution of $1024$.
        }
\label{code_comparison_RT}
\end{figure}                             
The first code is based on the use of pseudospectral methods to simulate
the RT instability. The spatial
discretization is Fourier in the periodic, horizontal directions and
Chebyshev in the gravitational direction. The temporal discretization is
3rd order Adams-Bashforth for the nonlinear terms and Crank-Nicholson for
the Laplacian terms. We use Werne's tau-correction scheme to achieve
incompressibility (\cite{JWerne_95}); the code is parallelized using MPI
(\cite{Dubey_99}),
and has been run on the Cray-T3E supercomputer at PSC.
The second code is based on the spectral element method
(\cite{HTufo_PFischer_99}).  It combines the spectral accuracy/efficiency
and the geometric flexibility of finite elements, and has been widely
applied to a variety of problems. 
%
%
%\begin{figure}[ht]
%%\epsfysize=3.5in
%%\epsfbox {Code_Comparison_RT.eps}
% \begin{center}
%  \begin{tabular}{c c}
%   {\psfig{file=Code_Comparison_RT.eps,height=2.6in,width=2.5in}} &
%   {\psfig{file=Spatial_res.eps,height=2.5in,width=2.5in}} \\
%   (a) & (b)
%  \end{tabular}
% \end{center}                   
%\caption{Panel (a): Comparison between the spectral-element code (solid line)
%	 and the spectral code (dashed line) for simulation of single
%         mode perturbation at the interface.  Plotted is the
%         mixing zone width $h$ as a function of time $\tau$.
%	 Panel (b): The spectral code shows exponential convergence for
%	 a fixed interface thickness of $0.15$ in the single
%	 mode simulation. The error
%	 is defined as the deviation from
%	 the simulation with a resolution of $1024$.     
%         }
%\label{code_comparison_RT}
%\end{figure}
%
Figure \ref{code_comparison_RT} (a) demonstrates
the satisfactory agreement between the two codes.  We plot
the penetration depth $h$ of the single mode perturbation as a function
of time $\tau$.  The solid line is from spectral-element code, and the
dashed line is from spectral code.  The spectral
resolution is $256$ by $512$ and the spectral element code
has an equival number of total grid points for this comparison.
The average difference in $h$ is as small as $10^{-4}$.
By comparing
results between spectral-element and spectral code, we have validated
the usage of spectral code in this particular situation, where
the small transitional region from one fluid
density to another may undermine the spectral convergence.                       
In addition to the code-code comparison, we have also
performed standard convergence
tests on the spectral code when applied to the miscible RT simulation 
to assure ourselves of the spectral convergence with the presence of 
a thin interface. We obtain the expected spectral
convergence if the interface is smooth and well resolved. 
The ideal
density discontinuity (or, in our case, a sharp temperature jump)
cannot be resolved in spectral codes without special treatment;
we therefore use a hyperbolic tangent or an error function for the
background temperature field to make sure that the initial temperature
field is smooth. From the smoothness of the interface, we obtain
spectral convergence as long as we resolve the interface. Figure 
\ref{code_comparison_RT} (b) shows the spatial convergence of the spectral
code when we fix the interface thickness and use single mode
perturbation as the initial condition.  
The convergence
test for the spectral element code is conducted separately for the
Orr-Sommerfeld equation as shown in table \ref{conv_table}.
The convergence is non-monotonic due to the fact that
the growth rates oscillate about the analytical
value. However, spectral convergence is clearly attained.

Having demonstrated the validation of the spectral code, we
present results from simulating the system with the spectral
code in the rest of the paper.
We adjust both
the resolution and the interface thickness so that there are enough grid
points across the background temperature profile.
\begin{table}
\begin{centering} % \vspace{1ex}
\begin{tabular}{ccccccccccc}
$N$&& $E(t_1)$   &&$error_1$ && $error_g$ \\ 
7  && 1.11498657 && 0.003963 && 0.313602 \\
9  && 1.11519192 && 0.003758 && 0.001820 \\
11 && 1.11910382 && 0.000153 && 0.004407 \\
13 && 1.11896714 && 0.000016 && 0.000097 \\
15 && 1.11895646 && 0.000006 && 0.000041 \\ 
\end{tabular} \\
\end{centering}
\caption{Spatial convergence of the spectral-element code
	 for the Orr-Sommerfeld problem: $K=15,\, dt=.003125$.
         $E(t_1)$ is the energy of the perturbation at
	 $t_1=25.1437$, the one period
	 of oscillation for the TS wave. 
	 $error_1$ is the error in energy when compared to theory,
	 and $error_g$ is the error in growth rate.
	 }             
\label{conv_table}
\end{table}

\section{``Random" perturbations at the interface: 2-D versus 3-D}

\begin{figure}[ht]
\begin{center}
\epsfysize=2.0in
\epsfbox{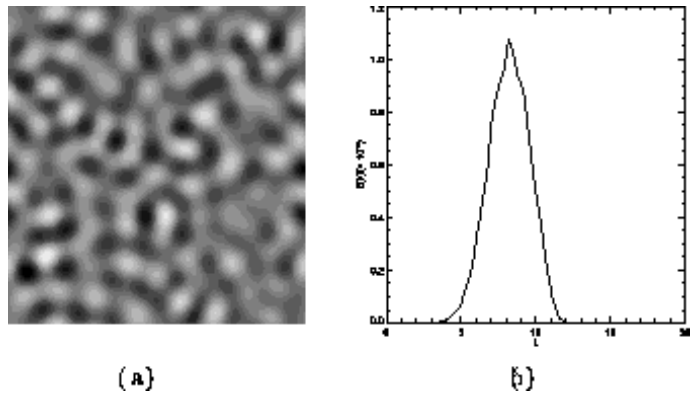}
\end{center}
% \begin{center}
%    \begin{tabular}{c c}
%    {\psfig{file=AC_init.eps,width=2.5in}} &
%    {\psfig{file=AC_init_spec.eps,width=2.5in}} \\
%    (a) & (b)
%    \end{tabular}
% \end{center}
\caption{Initial random perturbation at the interface $(z=0.06)$.
Panel (a) is the top view of the perturbation and
panel (b) is the power spectrum of the perturbation.}
\label{random_init}
\end{figure}

In this section, we focus on the consequence of perturbing the interface
with a spatially randomized disturbance. The initial
perturbation imposed at the interface in the 3-D simulation is shown in
figure \ref{random_init}. This perturbation is the outcome of
phase randomization of a spatial spectral distribution which peaks 
at $k\sim 8$ and whose bandwidth is approximately $4$ centered at the
peak. This initial perturbation has been defined as one of the canonical
test problems of the ``alpha group" collaboration (A. Cook \&
G. Dimonte, private
communication).  The perturbation spectrum (figure \ref{random_init} (b))
is designed so that the fastest growing mode
is well resolved for the given resolution
$(256\times 256\times 512)$.  The physical motivation for this
initial condition is that we want to have ``many" (for
our resolution, we have around $60$ small plumes at the
beginning), well resolved
small plumes seeded at the initial interface to study the
mixing of the instability in details. Later we
also show that the growth rate of this mode ($k\sim 8$) is
close to the growth rate 
for non-diffusive, inviscid  fluid (figure \ref{Early_Energetic3D}). 

We place the position of the initial interface at
$z=0.06$ ($-1\le z \le 1$), so that we can compare the behavior of
penetrating plumes as they approach the top and bottom walls near the end
of the simulations.\footnote{From our simulation, we conclude that the
effect of the top and bottom boundaries is not important to the
propagation of the mixing zone until the mixing zone width reaches 
$\sim 90\%$
of the box size.} The corresponding initial temperature profile used in
the 2-D simulations is taken from a planar cut through the 3-D initial
profile, e.g., the $y=0$ plane
$(T(x,y=0,z))$ (we also have conducted the 2-D simulations with other
planar cuts from the 3-D profile and found no material differences).
In all instances, we
treat the case $\nu = 10^{-3}$ and $\kappa=\nu$.

\subsection{Time evolution}
\subsubsection{Mixing zone width}
\begin{figure}[ht]
%\vspace{7.0in}
\begin{center}
\epsfysize=6.5in
\epsfbox{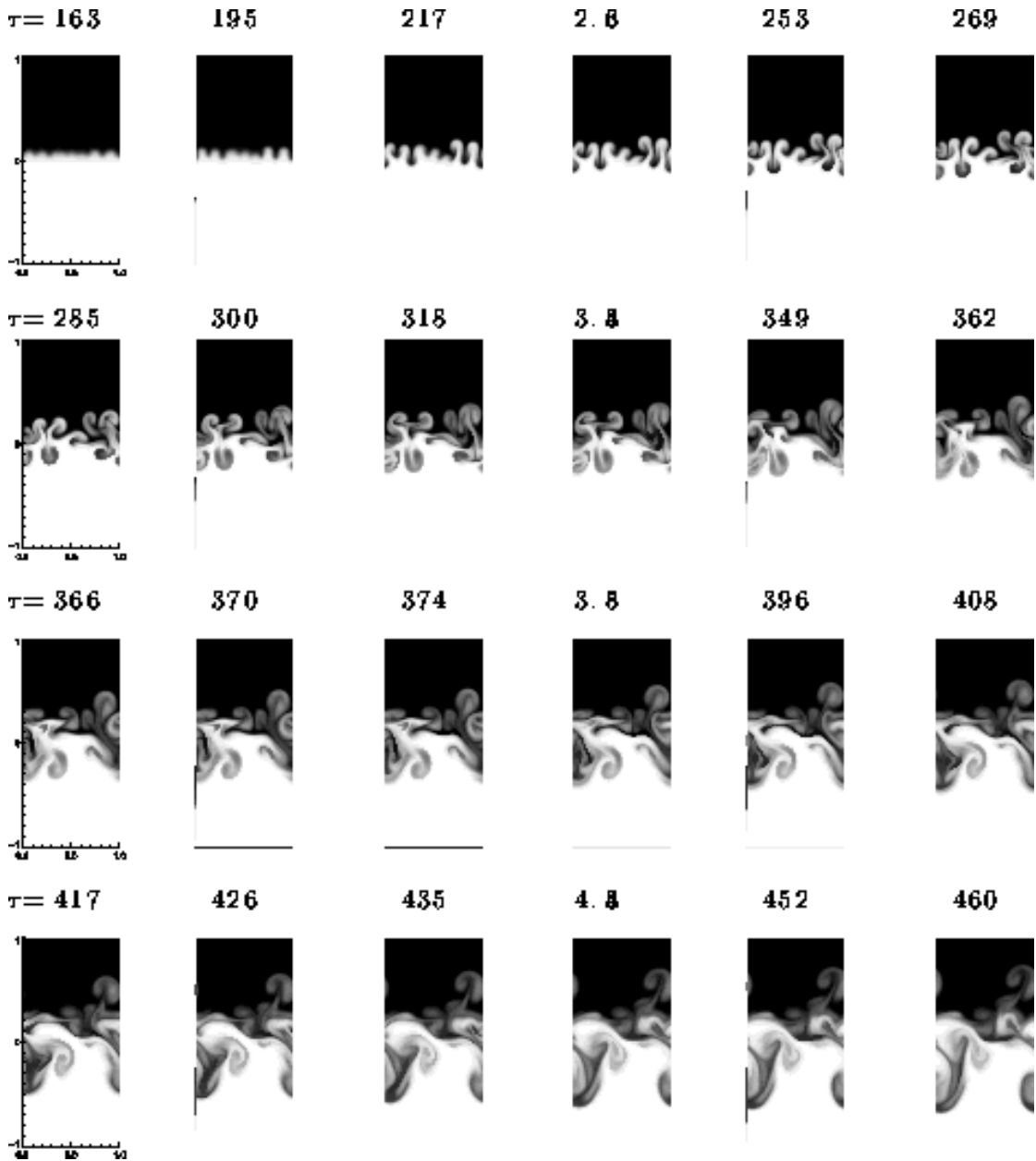}     
\end{center}
%\begin{center}
%\begin{tabular}{c c c c c c}
%$\tau=$1.63 & 1.95 & 2.17 & 2.36 & 2.53 & 2.69\\
%{\psfig{file=2Drun_yslice01.eps,width=1.0in}} &
%{\psfig{file=2Drun_yslice02.eps,width=1.0in}} &
%{\psfig{file=2Drun_yslice03.eps,width=1.0in}} &
%{\psfig{file=2Drun_yslice04.eps,width=1.0in}} &
%{\psfig{file=2Drun_yslice05.eps,width=1.0in}} &
%{\psfig{file=2Drun_yslice06.eps,width=1.0in}} \\
%$\tau=$2.85 & 3.00 & 3.18 & 3.34 & 3.49 & 3.62\\
%{\psfig{file=2Drun_yslice07.eps,width=1.0in}} &
%{\psfig{file=2Drun_yslice08.eps,width=1.0in}} &
%{\psfig{file=2Drun_yslice09.eps,width=1.0in}} &
%{\psfig{file=2Drun_yslice10.eps,width=1.0in}} &
%{\psfig{file=2Drun_yslice11.eps,width=1.0in}} &
%{\psfig{file=2Drun_yslice12.eps,width=1.0in}} \\
%$\tau=$3.66 & 3.70 & 3.74 & 3.85 & 3.96 & 4.08\\
%{\psfig{file=2Drun_yslice13.eps,width=1.0in}} &
%{\psfig{file=2Drun_yslice14.eps,width=1.0in}} &
%{\psfig{file=2Drun_yslice15.eps,width=1.0in}} &
%{\psfig{file=2Drun_yslice16.eps,width=1.0in}} &
%{\psfig{file=2Drun_yslice17.eps,width=1.0in}} &
%{\psfig{file=2Drun_yslice18.eps,width=1.0in}} \\
%$\tau=$4.17 & 4.26 & 4.35 & 4.43 & 4.52 & 4.60\\
%{\psfig{file=2Drun_yslice19.eps,width=1.0in}} &
%{\psfig{file=2Drun_yslice20.eps,width=1.0in}} &
%{\psfig{file=2Drun_yslice21.eps,width=1.0in}} &
%{\psfig{file=2Drun_yslice22.eps,width=1.0in}} &
%{\psfig{file=2Drun_yslice23.eps,width=1.0in}} &
%{\psfig{file=2Drun_yslice24.eps,width=1.0in}} \\
%\end{tabular}
%\end{center}
\caption{Time series of the temperature field in a $1(x)$ by $2(z)$ box
	 from the 2D simulations.
	  The interface between cold (black) and warm (white)
   fluids is perturbed by the $\delta T(x,y=0)$ slice taken from
   the spatially random disturbances in the 3D simulations.}
\label{F2Dtemp}
\end{figure}

\begin{figure}[ht]
%\vspace{7.0in}
\begin{center}
\epsfysize=6.5in
\epsfbox{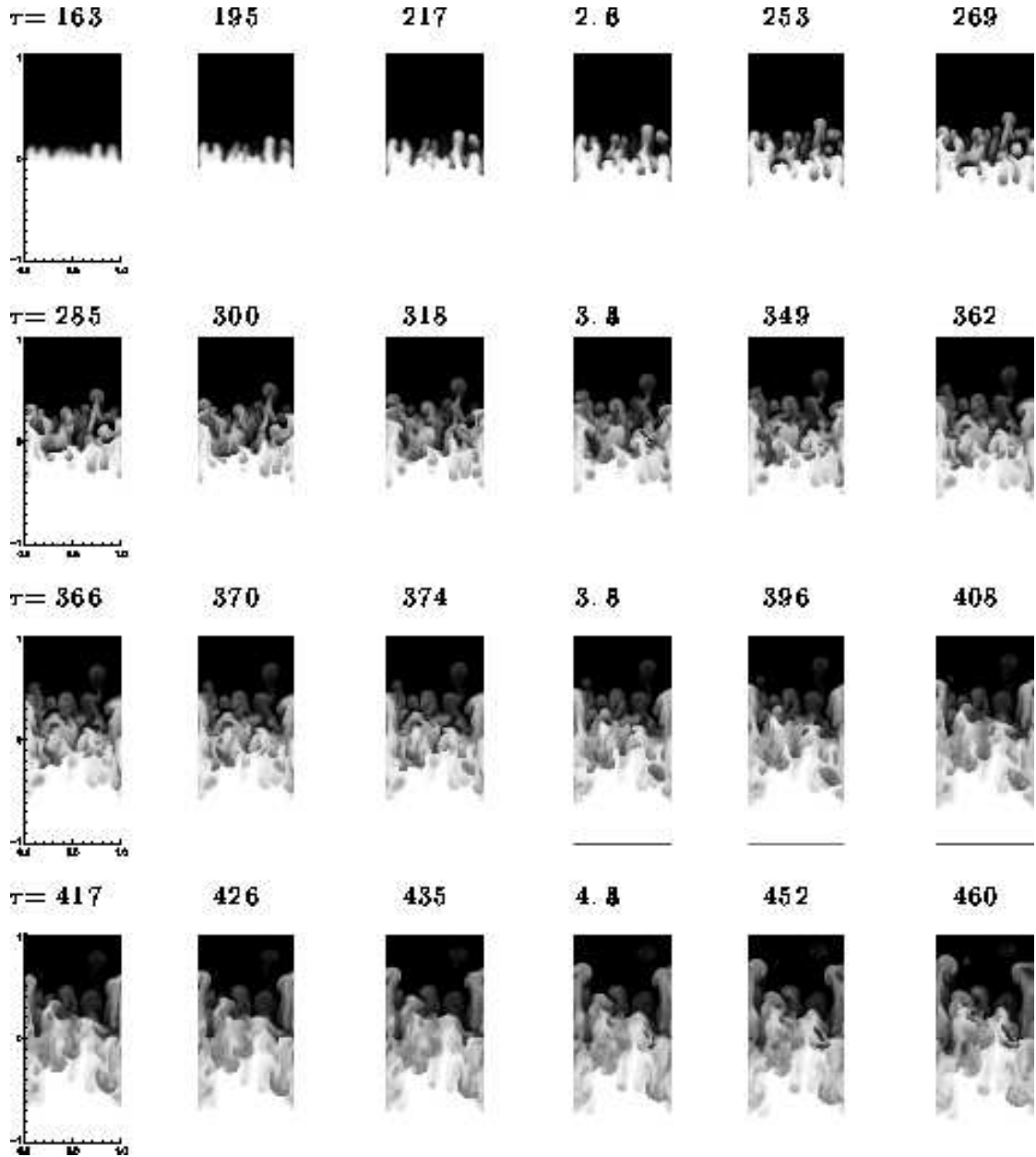}     
\end{center}
%\begin{center}
%\begin{tabular}{c c c c c c}
%$\tau=$ 1.63 & 1.95 & 2.17 & 2.36 & 2.53 & 2.69 \\
%{\psfig{file=3Drun_yslice01.eps,width=1.0in}} &
%{\psfig{file=3Drun_yslice02.eps,width=1.0in}} &
%{\psfig{file=3Drun_yslice03.eps,width=1.0in}} &
%{\psfig{file=3Drun_yslice04.eps,width=1.0in}} &
%{\psfig{file=3Drun_yslice05.eps,width=1.0in}} &
%{\psfig{file=3Drun_yslice06.eps,width=1.0in}} \\
%$\tau=$ 2.85 & 3.00 & 3.18 & 3.34 & 3.49 & 3.62 \\
%{\psfig{file=3Drun_yslice07.eps,width=1.0in}}&
%{\psfig{file=3Drun_yslice08.eps,width=1.0in}} &
%{\psfig{file=3Drun_yslice09.eps,width=1.0in}} &
%{\psfig{file=3Drun_yslice10.eps,width=1.0in}} &
%{\psfig{file=3Drun_yslice11.eps,width=1.0in}} &
%{\psfig{file=3Drun_yslice12.eps,width=1.0in}} \\
%$\tau=$ 3.66 & 3.70 & 3.74 & 3.85 & 3.96 & 4.08 \\
%{\psfig{file=3Drun_yslice13.eps,width=1.0in}} &
%{\psfig{file=3Drun_yslice14.eps,width=1.0in}} &
%{\psfig{file=3Drun_yslice15.eps,width=1.0in}} &
%{\psfig{file=3Drun_yslice16.eps,width=1.0in}} &
%{\psfig{file=3Drun_yslice17.eps,width=1.0in}} &
%{\psfig{file=3Drun_yslice18.eps,width=1.0in}} \\
%$\tau=$ 4.17 & 4.26 & 4.35 & 4.43 & 4.52 & 4.60 \\
%{\psfig{file=3Drun_yslice19.eps,width=1.0in}} &
%{\psfig{file=3Drun_yslice20.eps,width=1.0in}} &
%{\psfig{file=3Drun_yslice21.eps,width=1.0in}} &
%{\psfig{file=3Drun_yslice22.eps,width=1.0in}} &
%{\psfig{file=3Drun_yslice23.eps,width=1.0in}} &
%{\psfig{file=3Drun_yslice24.eps,width=1.0in}} \\
%
%\end{tabular}
%\end{center}
\caption{Time series of the temperature from the 3D simulations.
	 The slices $(T(x,y=0,z))$ are taken at the same times as
  those for the corresponding 2D frames (figure \ref{F2Dtemp}). }
\label{F3Dtemp}
\end{figure}

\begin{figure}[htbp]
%\vspace{15pc}
\begin{center}
\epsfysize=2.0in
\epsfbox{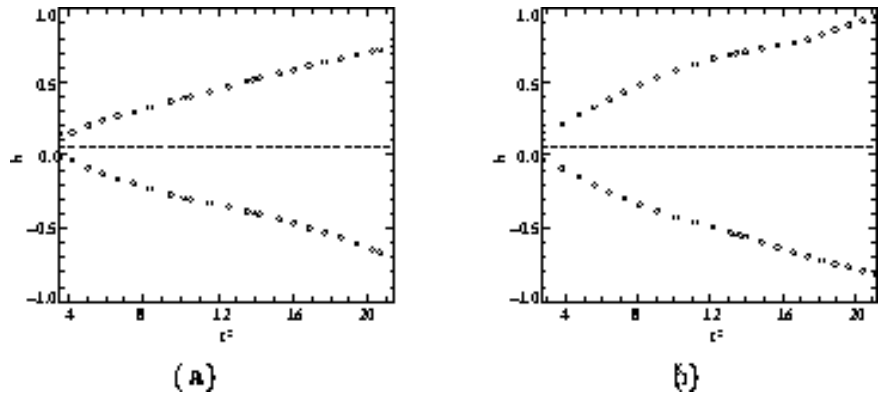}     
\end{center}
%\begin{center}
%\begin{tabular}{c c}
%{\psfig{file=cs128erf_2d_h01_a01.eps,width=2.7in}} &
%{\psfig{file=cs128erf_3d_h01_a01.eps,width=2.7in}} \\
%(a) & (b) \\
%
%\end{tabular}
%\end{center}
\caption{Penetration depths from 2D (a) and 3D (b) simulations for
	 the random perturbation case.  For the ascending structures,
         $h$ is defined as the maximum value where the temperature
         reaches above $99\%$ of the cold fluid temperature. For
         the descending mixing zone boundary, $h$ is defined as the minimum
         value where the temperature drops below $99\%$ of the hot fluid
         temperature.}
\label{Penetration2V3}
\end{figure}

We first compare the 2-D and 3-D temporal evolution;
the panels in figure
\ref{F2Dtemp} show the 2-D time evolution of the temperature,
while figure \ref{F3Dtemp} shows the time evolution of $y=0$ slices from
the 3-D simulations.           
Figure \ref{Penetration2V3} provides
the penetration depths as functions of
dimensionless time for 2-D and 3-D. (To scale
our time according to Youngs' time scale (\cite{Youngs91}), we multiply $t$
with $(\beta g/l )^{1/2}$; thus $\tau \equiv (\beta g/l )^{1/2}
t$, where $l$ is the half height of the box).
The penetration depth is determined by measuring the distance
between boundaries of volume fraction of $1\%$ of the hot fluid
and $99\%$ of the cold fluid.                                      
\clearpage

\clearpage
%
%Figure \ref{Penetration2V3} provides
%the penetration depths as functions of
%dimensionless time for 2-D and 3-D. (To scale
%our time according to Youngs' time scale (\cite{Youngs91}), we multiply $t$
%with $(\beta g/l )^{1/2}$; thus $\tau \equiv (\beta g/l )^{1/2}
%t$, where $l$ is the half height of the box).
%The penetration depth is determined by measuring the distance
%between boundaries of volume fraction of $1\%$ of the hot fluid
%and $99\%$ of the cold fluid.     
%

\begin{figure}[ht]
%\vspace{15pc}
\begin{center}
\epsfysize=2.0in
\epsfbox{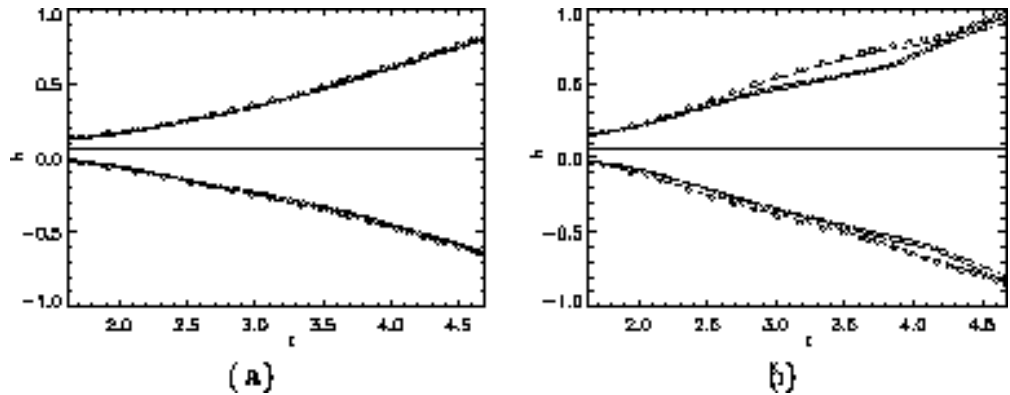}     
\end{center}
%\begin{center}
%\begin{tabular}{c c}
%{\psfig{file=s128erf_h_total_2d_02a.eps,width=2.7in}} &
%{\psfig{file=s128erf_h_total_02a.eps,width=2.7in}} \\
%(a) & (b) \\
%
%\end{tabular}
%\end{center}
\caption{Penetration depths from 2D (a) and 3D (b) for different
	 volume cuts.  The 2D curves are from the ensemble average
	  and show similar evolution
   as in figure \ref{Penetration2V3}(a).
   The diamonds are for $0.5-99.5\%$ volume
   cuts of hot-cold fluids, the dashed lines are for
   $1-99\%$ volume cuts, the dash-dotted lines are for
   $3-97\%$ volume cuts, and the dash-dot-dotted lines are
   for $5-95\%$.
   }
\label{volume_cuts}
\end{figure}

In figure \ref{volume_cuts} we
show how $h$ varies depending on different volume cuts of hot and
cold fluids, and we conclude that penetration depths are well defined
if the volume cuts are narrower than $1-99\%$ range, which is what
we have adopted to calculate results presented below.
We plot the penetration depths against
$\tau^2$ instead of $\tau$ in order to clarify the scaling behavior. For
both 2-D and 3-D, $h\propto  \alpha \tau^2$
during early stage of the evolution for $1.8\le \tau \le 4.4$.
The $\alpha$ for this period of time is $\sim 0.017$ for the 2-D case,
and $\alpha \sim 0.03$ for the 3-D case. Thus, the 3-D mixing zone
broadens much faster than its 2-D counterpart.                       

We also observe that the
scaling range $4\le \tau^2 \le 12$ is twice the range found in
Youngs' work. The mixing zone fills the entire tank at around $\tau = 4.7$
in our case, compared to the values of $2.9\sim 3.1$ from simulations with
low Mach numbers and high Atwood numbers in \cite{Youngs91}. Finally, we
observe that at time $\tau^2 \sim 12$, the increasing
$h$ levels off; $h$ then
resumes $\tau^2$ scaling at around $\tau^2 \sim 16$; 
we will discuss this phenomenon in detail in the following
subsection.  We further remark that the alpha coefficient for both
2 and 3-D presented above is only for early evolution.  

\subsubsection{2-D flows versus 3-D flows: Finger geometry}
\begin{figure}[ht]
%\vspace{15pc}
\begin{center}
\epsfysize=2.0in
\epsfbox{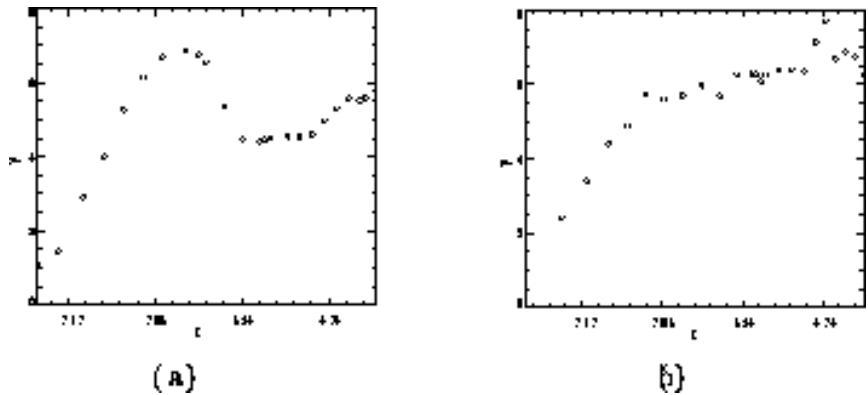}     
\end{center}
%\begin{center}
%\begin{tabular}{c c}
%{\psfig{file=c2Drun_y_arclength_a01.eps,width=2.7in}} &
%{\psfig{file=cs128erf_2dy_arclength_a01.eps,width=2.7in}} \\
%(a) & (b) \\
%\end{tabular}
%\end{center}
\caption{Curvelengths of the isothermal contour $(T=0.0)$
for both 2D (a) and
3D (b).}
\label{Curvelength2V3}
\end{figure}

In order to understand these results, we first note that the structure of
2-D fingers (plumes) is horizontally stretched, whereas in 3-D, fingers remain
vertical until they merge and form bigger plumes. This implies that the
2-D flow has a significantly larger horizontal 
velocity component when compared to
3-D; we return to this observation later when we discuss the evolution
of single plumes. In both 2-D and 3-D, we find one large thermal pushing
the envelopes at late times; however, in 3-D, we observe more local,
small-scale structures in the interior of the mixing region than in the
2-D case. The large plumes push the edges of the mixing region, and
dominate the expansion of the mixing zone at late times. The interior of
the mixing zone is also very different between 2-D and 3-D: In 2-D, the
interior structure is dominated by the large plumes, and therefore the
general structure of the interior is simple. In 3-D, the interior is a
combination of boundaries between plumes, residuals of mergers, and the
wake behind the faster plumes; the interior of the 3-D mixing region is
thus much more complicated than that of the 2-D mixing zone. 
%(Because
%there are more local structures in 3-D than in 2-D, one concludes  that
%mixing in 3-D is more efficient than in 2-D.)\ \ 
A simple quantitative
measure of this difference in structural complexity between 2 and 3-D can
be constructed by measuring the stretching of isothermal contours during
the course of R-T instability. Thus, figure
\ref{Curvelength2V3} shows the time variation of the isothermal contour
length (area) corresponding to $T=0$ for 2-D (3-D). 
The isothermal contour length is calculated as follows: we
first identify points on the isothermal curves, we then calculate the
length along the curve by linear interpolation between points.  Similar
technique has been utilized to calculate the iso-surface in 3D.

Initially, the 2-D
arclength increases much faster than the corresponding 3-D
arclength; this is because the 2-D
isothermal contours are subject to horizontal stretching comparable to
their vertical stretching. At late times, however, the 2-D isothermal
contour breaks off into several connected contours, and thus a sharp
decrease in the arc length results. In the 3-D case, the arc length
increases almost monotonically until a maximum value is reached.
This peak is due to the stretching by the shear flow as the edge
plumes push the envelopes sideways. The break-off of this isolevel near
the front explains the decrease in the arc length at $\tau\sim 4.2$.

\subsubsection{Energetics: 2-D versus 3-D}
\begin{figure}[ht]
%\vspace{15pc}
\begin{center}
\epsfysize=2.0in
\epsfbox{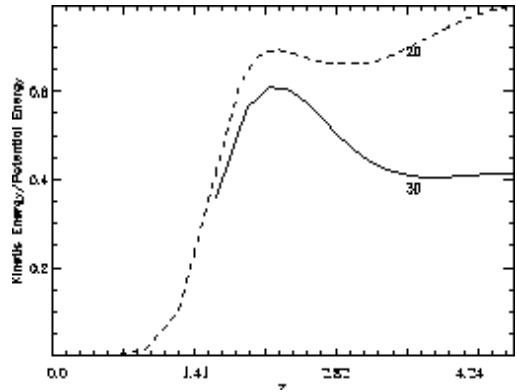}     
\end{center}
%\begin{center}
%\epsfysize=3.0in
%\epsfbox {c3D_2D_energetics_b.eps}
%\end{center}
\caption{
	Ratio of kinetic energy to gravitational potential energy: 2D versus 3D.
         The kinetic energy is the volume integral of the kinetic
         energy density, and the potential energy is the volume integral
         of the potential energy available in the system. For $\tau \le 1.4$,
         the growth is exponential and is similar for both 2 and 3D, after this
	 period, 2-D motions are much more efficient in extracting potential
	 energy than 3-D motions.}
\label{Energetics2V3}
\end{figure}

We now focus on the comparison between 2-D and 3-D energetics: We first
plot the ratio of kinetic energy to potential energy available to the
flow  as a function of time in figure \ref{Energetics2V3}. 
Our results show a remarkable feature: the 2-D nonlinear
Rayleigh-Taylor instability is more efficient at extracting 
gravitational potential energy than its 3-D counterpart, yet
(as shown in \S 3.1), the 2-D mixing zone grows more slowly
than its 3-D counterpart.  As we will show below, this 
apparent contradiction is resolved when one looks at how the
kinetic energy is partitioned.  
In \cite{Youngs91}, Youngs found
the two-dimensional flow is less dissipative than
the three-dimensional flow.  In his simulations he found that
the kinetic energy reaches a maximum at $\tau\sim 6$ for both 
two and three dimensions.  However, the 3-D flow is more ``dissipative"
and therefore the 3-D kinetic energy decreases significantly from the
maximum after $\tau=6$.  He also showed that in 3-D the difference
between potential energy and kinetic energy 
is greater than in 2-D.  This is consistent
with our findings.  Later we show the energy spectrum of the flow
inside the mixing zone from our 3-D 
simulations (figure \ref{Ek_center_MZ}).  
Similar energy spectrum can also be found in \cite{Youngs91}.

\begin{figure}[ht]
%\vspace{15pc}
\begin{center}
\epsfysize=2.0in
\epsfbox{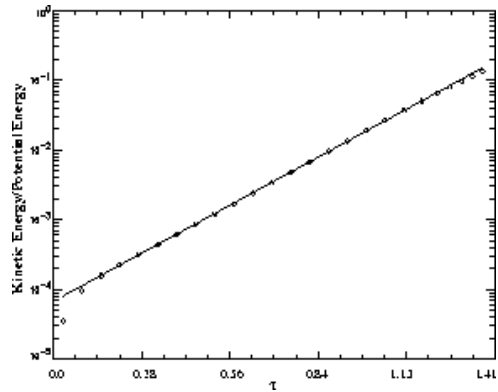}     
\end{center}
%\begin{center}
%\epsfysize=3.5in
%\epsfbox {cearly_energetic_02.eps}
\caption{Early evolution of the 3D random perturbation. The diamonds are
	 data from the numerical simulation.  The solid line is the best
  fit on the log-linear plot (for $0.28\le \tau \le 1.13$),
  and the duration of the linear growth
  is about $4$ e-folding time with a grow rate of $8$.
  The growth rate obtained from the
  analytic results ($n^2=gAk$, with $gA=10$ and $k\sim 8$ from the
  power spectrum of our initial perturbation at the interface), is
  $9$.}
\label{Early_Energetic3D}
%\end{center}
\end{figure}

In figure
\ref{Early_Energetic3D} we plot this energy ratio for the 3-D case at
early times (the 2-D growth is almost the same). 
The growth rate from
figure \ref{Early_Energetic3D} is $\sim 8$, close to the growth rate
($\sim 9$) for $k=8$ (where the spatial perturbation spectrum peaks)
from the analytic results for the
inviscid case (\cite{Chandra}). Thus one concludes that the $k=8$ mode
dominates the evolution of the initially randomly seeded perturbation. We
also note that at around $\tau \sim 1.3$, which is $6$ e-folding times
after the initiation of linear growth at $\tau \sim 0.3$, we begin to
observe saturation due to the nonlinear terms; this marks the
beginning of the nonlinear evolutionary stage.  After this linear regime,
the peak of the spectrum decreases in amplitude and moves toward the
small wavenumber end, an indication of formation of large scales.

\subsection{Average Quantities and the spatial structure of the mixing zone}
\subsubsection{Average Quantities}
Here we focus on the evolution of the horizontally averaged
quantities: the horizontal components of the vorticities and the advective
thermal fluxes $\vec{u}T$; after horizontal averaging, each of these
quantities is only a function of the vertical coordinate $(z)$ and time
$(\tau)$. We note that the buoyancy flux is the vertical component of the
advective flux, while the buoyancy gradient is the vertical component of
the advective thermal flux.

\begin{figure}[ht]
%\vspace{15pc}
\begin{center}
\epsfysize=2.0in
\epsfbox{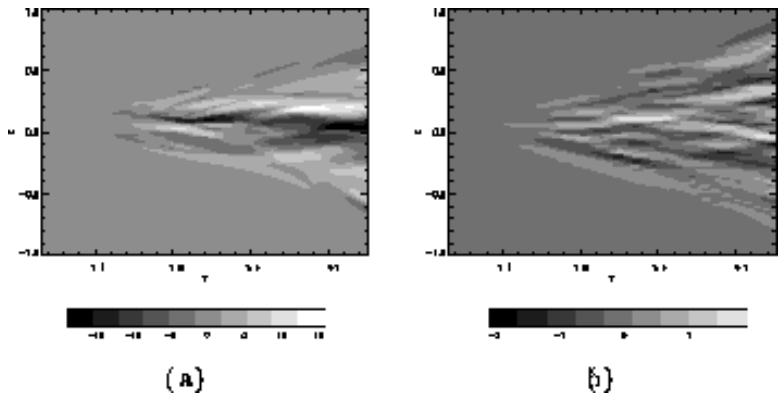}     
\end{center}
%\begin{center}
%\begin{tabular}{c c}
%{\psfig{file=c2D_vty01.eps,width=2.5in}} &
%{\psfig{file=c3D_hvt01.eps,width=2.5in}} \\
%{\psfig{file=2D_vty01_bar.eps,width=2.0in}} &
%{\psfig{file=3D_hvt01_bar.eps,width=2.0in}}\\
%(a)&(b)\\
%\end{tabular}
%\end{center}
\caption{Horizontally averaged vorticity as functions of $z$ and $t$.
(a) is the 2D vorticity,
(b) is the mean horizontal 3D vorticities.}
\label{HVty2V3}
\end{figure}

First, we concentrate on the average horizontal components of the
vorticity (figure \ref{HVty2V3}). The average of the horizontal vorticity
is basically the vertical gradients of the horizontal components of the
velocity, as we adopt periodic boundary conditions in the lateral
directions. We note that the amplitude of the 2-D vorticity is
almost $5$ times larger than that of the 3-D vorticity components.  The
evolution of the 2-D vorticity (panel (a)) is indicative of the strong
shear strength at the center of the mixing region. The patches of
interchanging shades of gray in the time-space plot of
the 3-D vorticity components indicate the merging events within the mixing
region. We observe a similar pattern in the plot of 2-D vorticity, but
only at early times; at late times, the 2-D space-time plot is more
indicative of the roll-up of big thermals.                   

\begin{figure}[ht]
%\vspace{15pc}
\begin{center}
\epsfysize=2.0in
\epsfbox{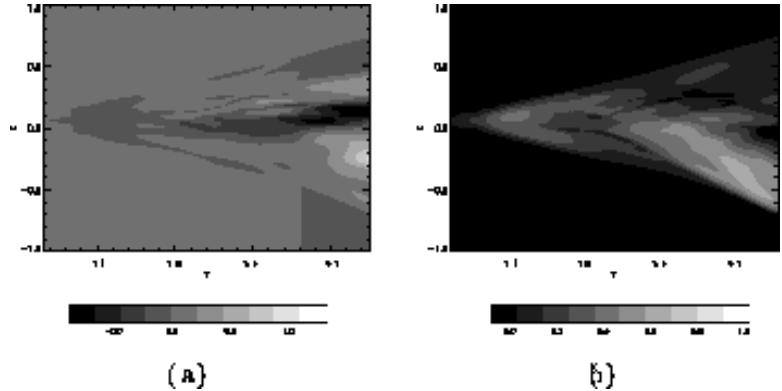}     
\end{center}
%\begin{center}
%\begin{tabular}{c c }
%{\psfig{file=c2D_ut01_a.eps,width=2.5in}} &
%{\psfig{file=c2D_wt01_a.eps,width=2.5in}} \\
%{\psfig{file=2D_ut01_bar_a.eps,width=2.5in}} &
%{\psfig{file=2D_wt01_bar_a.eps,width=2.5in}} \\
%(a) & (b) \\
%
%\end{tabular}
%\end{center}
\caption{Average advective fluxes (2D) as
functions of $z$ and $t$. (a) is the
horizontal flux and (b) is the vertical (buoyancy) flux.}
\label{ATF_2D}
\end{figure}

\begin{figure}[ht]
%\vspace{15pc}
\begin{center}
\epsfysize=2.0in
\epsfbox{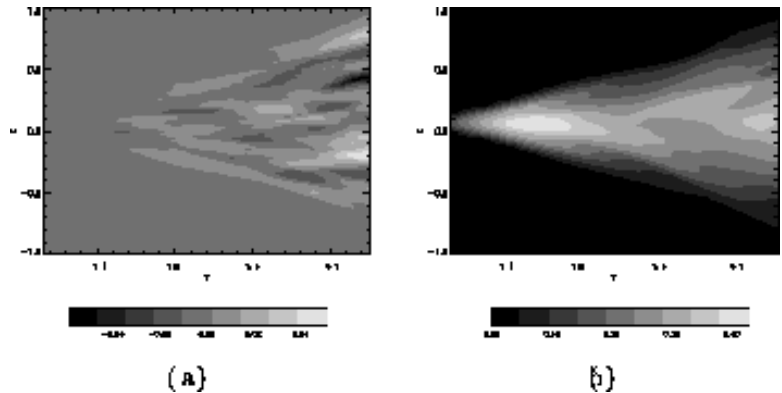}     
\end{center}
%\begin{center}
%\begin{tabular}{c c}
%{\psfig{file=c3D_ahtf01.eps,width=2.5in}} &
%{\psfig{file=c3D_avtf01.eps,width=2.5in}} \\
%{\psfig{file=3D_ahtf01_bar.eps,width=2.0in}} &
%{\psfig{file=3D_avtf01_bar.eps,width=2.0in}} \\
%(a)&(b)\\
%\end{tabular}
%\end{center}
\caption{Horizontally averaged advective fluxes (3D)
	 as functions of $z$ and $t$. (a)
         is the mean horizontal components and (b)
         is the vertical (buoyancy) flux.
}
\label{ATF_3D}
\end{figure}

In figure \ref{ATF_2D} we show the time-space
plot of the average advective
thermal fluxes in the horizontal and vertical directions;
the corresponding plots for 3-D are shown in figure
\ref{ATF_3D}. We note that in 2-D, the two fluxes are comparable in
amplitude, whereas in 3-D, the horizontal advective fluxes are of
comparable amplitude yet are both smaller than the vertical flux
(buoyancy flux) by a factor of $5$. This is a manifestation of the fact
that in 3-D, the flow is dominantly in the preferred (i.e., vertical)
direction, while the corresponding flow in 2-D is strongly
sheared (and has comparable magnitude in the horizontal and vertical
directions). The space-time plots for the diffusive thermal fluxes
basically reveal the same broadening structure of the mixing zone.
The horizontal components in 2-D and 3-D are comparable,
and the 2-D vertical diffusive flux is larger than the 3-D vertical flux,
as found in the average advective fluxes.

To briefly summarize: We observe stronger 2-D vorticity
generation compared to 3-D. Consequently, we observe that the 2-D
horizontal transport is much more effective than 3-D. The relative
efficiency of vertical advective transport to horizontal transport,
in contrast, is much higher in 3-D than that in 2-D.

\subsubsection{Turbulence and mixing within the Mixing Zone}
\begin{figure}[ht]
%\vspace{15pc}
\begin{center}
\epsfysize=2.0in
\epsfbox{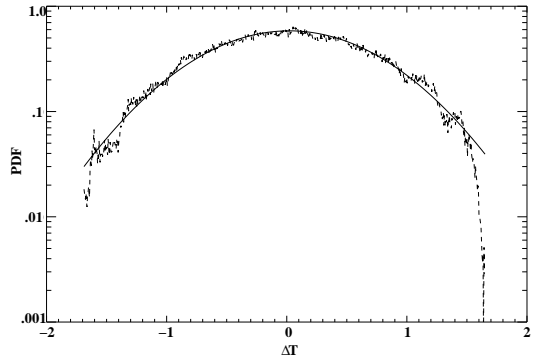}     
\end{center}
%\begin{center}
%\begin{tabular}{c}
%{\psfig{file=PDF_ieq201.eps,width=4.0in}}
%\end{tabular}
%\end{center}
\caption{PDF of the temperature fluctuation
	$(\delta T=T-T_0)$ within the mixing zone
	$(\Delta z=0.025)$ right at the position of the original interface
	$(z=0.06)$.}
\label{PDF_MZ01}
\end{figure}

\begin{figure}[ht]
%\vspace{15pc}
\begin{center}
\epsfysize=2.0in
\epsfbox{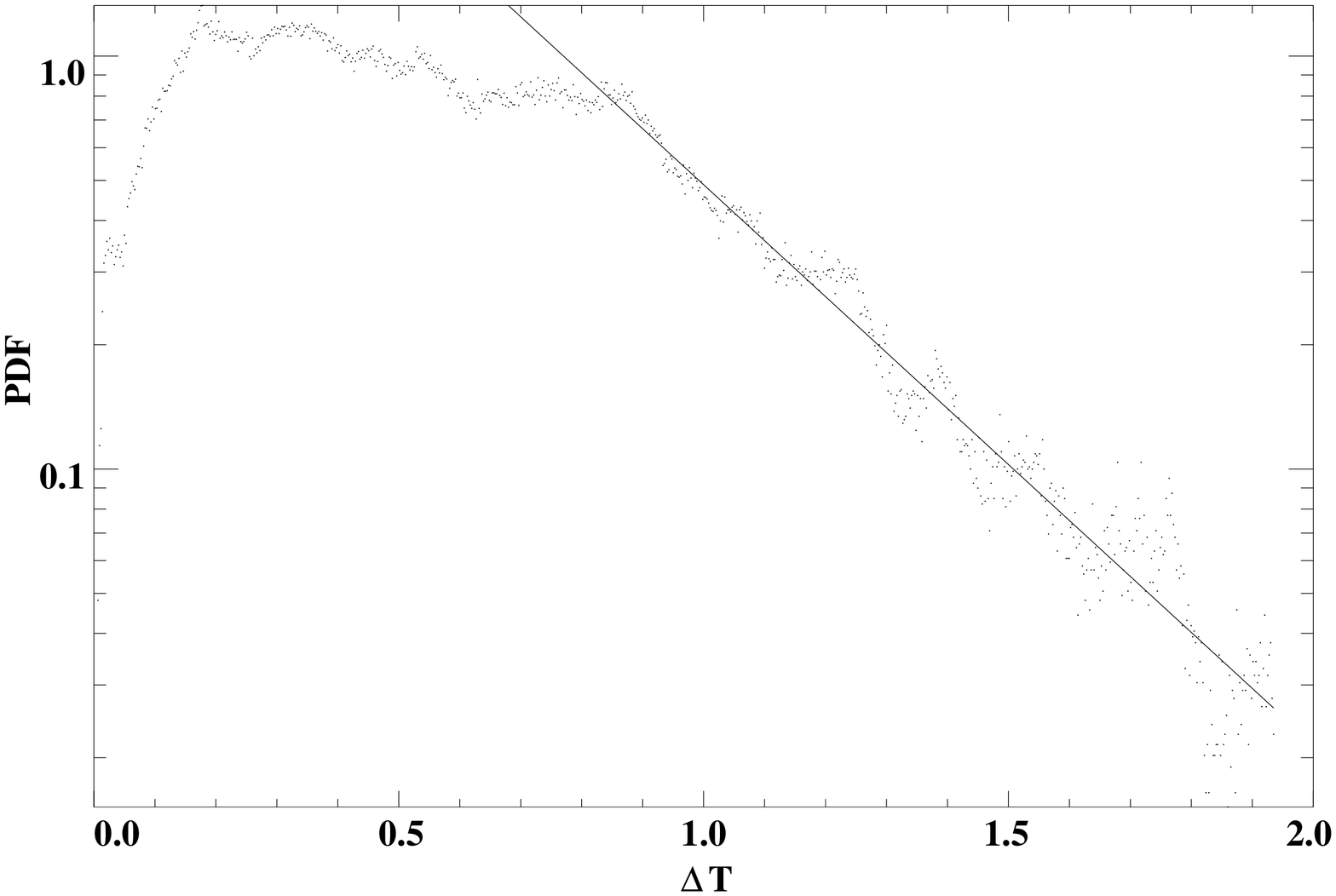}     
\end{center}
%\begin{center}
%\begin{tabular}{c}
%{\psfig{file=PDF_tmp03_7001.eps,width=4.0in}}
%\end{tabular}
%\end{center}
\caption{PDF of the temperature fluctuation
	 $(\delta T=T-T_0)$ within the mixing zone
         $(\Delta z=0.025)$ at $z=0.21$.}
\label{PDF_MZ05}
\end{figure}

\begin{figure}[htbp]
%\vspace{15pc}
\begin{center}
\epsfysize=2.0in
\epsfbox{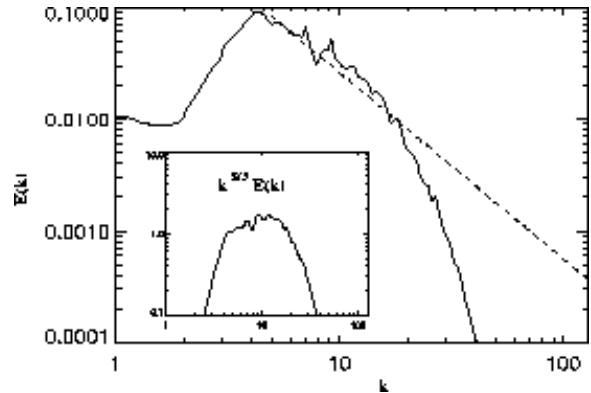}     
\end{center}
%\begin{center}
%\begin{tabular}{c}
%{\psfig{file=s128erf_spec_170_00a.eps,width=4.0in}}
%\end{tabular}
%\end{center}
\caption{Energy spectrum of the flow at the end of the simulatoin.
	 The dashed line is the Kolmogorov $-5/3$ scaling.  The inset
         shows that the inertial range is from $k\sim 4$ to $k\sim 20$.}
\label{Ek_center_MZ}
\end{figure}

As shown in the time series for both 2 and 3-D, the spatial structure of
the 2-D mixing zone is relatively much simpler than 3-D: the 2-D mixing
zone is dominated by one or two big plumes while in 3-D the center of the
mixing zone is full of small scale structures. This is clearly manifested
by the temperature fluctuation probability 
distribution function (PDF) within a thin horizontal slice
of the 3-D mixing zone. In our analysis we find a thickness of $\Delta
z=0.06$ (the initial interfacial thickness) suffices to provide good
statistics for the PDF. If we place this thin slice at the position of
the initial interface, we find the PDF of the temperature fluctuations
($\Delta T \equiv T-T_0$) to be a gaussian
centered at $\Delta T=0$ (figure \ref{PDF_MZ01}). 
As we move this thin slice away from the
initial interface, the PDF transitions from a gaussian distribution to an
exponential distribution (figure \ref{PDF_MZ05}). 
Moving further than $1/3$ of the mixing zone
width from the center, the PDF of the thin slice transitions again from an
exponential distribution to a yet steeper distribution 
(more like a delta function
centered at either $-2$ or $2$). 
The guassianity in the scalar PDF at the center of the mixing zone 
(figure \ref{PDF_MZ01}) implies turbulent mixing of the scalar
near the center.
For the flow inside the mixing zone, the direct evidence for turbulence is
a $-5/3$ Kolmogorov energy spectrum.
Due to our numerical resolution (256 by 256 by 512) the scaling
range is small as shown in figure \ref{Ek_center_MZ}, where 
the energy spectrum shows an
inertial range from $k\sim 4$ to $k\sim 18$.  
\noindent

\noindent

As we move from the center of the mixing
towards the boundary, 
the exponential PDF of the scalar at $z=0.5$ implies that the
flow that mixes the scalar is dominated by large scale.  
This indicates that, for the miscible RT
instability and the parameters that we explore, 
turbulent mixing exists only near the
initial interface, and that most of the mixing zone is still dominated by
finger structures. These finger structures (plumes) near the edge are
responsible for the broadening of the mixing zone. 
As we will show in \S
5, the propagation speed of a plume in an ambient environment is directly
proportional to its circulation. Also shown in \S 5 is that, if the drag
is small, these plumes initiated by the perturbation at the RT interface
will be free falling (rising) without any turbulent scaling involved.
Furthermore, if the plumes pushing the edge of the mixing zone experience
a different flow in the background, we would expect  a different mixing
zone broadening. For example, if the horizontal components of the flow
are comparable to the vertical component, we expect the propagation of
the mixing zone to slow down. Therefore, the transition of the ambient
background flows would serve as an indication of deviation from the free
falling phase of the mixing zone. In the next subsection we will provide
evidence for alteration of flow patterns inside the mixing zone in the
3-D case.   Combining all the results, we are able to demonstrate the
correlation between the transition of the interior flows and the
transition of the mixing zone broadening.

\subsection{Mixing Zone Broadening and the Internal Flow Structures}
\begin{figure}
%\vspace{15pc}
\begin{center}
\epsfysize=2.0in
\epsfbox{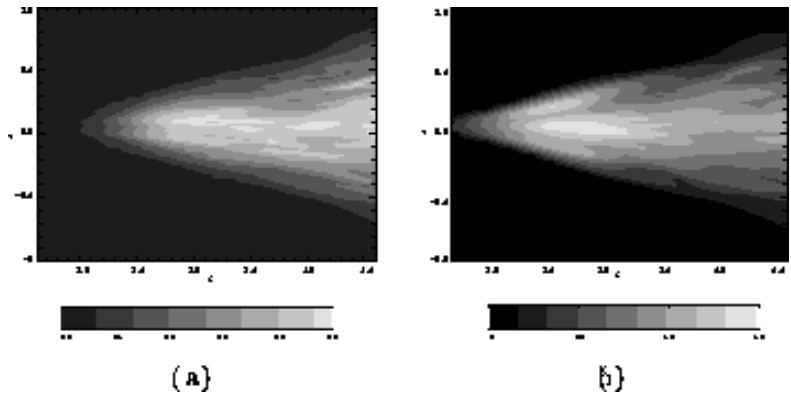}     
\end{center}
%\begin{center}
%\begin{tabular}{c c}
%{\psfig{file=cDiss_time-space_ho01.eps,width=2.5in}} &
%{\psfig{file=cDiss_time-space_0301.eps,width=2.5in}} \\
%
%{\psfig{file=cDiss_time-space_ho_box01.eps,width=2.5in}} &
%{\psfig{file=cDiss_time-space_03_box01.eps,width=2.5in}} \\
%(a)&(b) \\
%
%\end{tabular}
%\end{center}
\caption{
      Average viscous dissipation rates as functions of $z$ and $t$.
      (a) is the dissipation rate due to the horizontal components of the flow,
      and (b) is the vertical component.}
\label{Diss_3D}
\end{figure}

\begin{figure}
%\vspace{15pc}
\begin{center}
\epsfysize=2.0in
\epsfbox{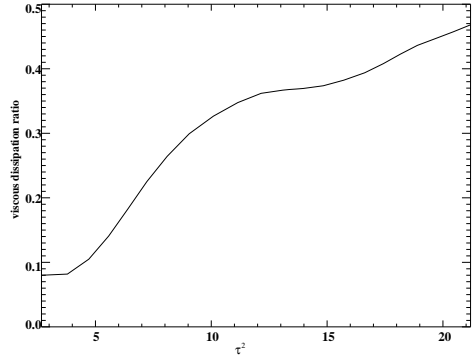}     
\end{center}
%\epsfysize=3.0in
%\epsfbox {cDiss_ratio_ho_0301.eps}
%\end{center}
\caption{Ratio of the volume integrated
	 viscous dissipation rates as a function of time $(\tau)$:
         the horizontal component of the dissipation rate
         to the vertical component.}
\label{Diss_ratio}
\end{figure}

We now focus on the relationship between the propagation of the mixing zone
and the internal flow structures 
at around $\tau^2 = 10$ in the 3-D case. In figure \ref{Diss_3D} we show
the average horizontal and vertical components of viscous dissipation
rate: the horizontal component is simply $\nu (u_1\nabla^2 u_1 +
u_2\nabla^2 u_2)/2$ and the vertical component is $\nu(u_3\nabla^2 u_3)$.
In figure \ref{Diss_ratio} we show the ratio of the volume integral of
these two components. 

Some observations from these results can help
us gain insight into the mixing zone evolution:
we note that average buoyancy flux (panel (b) in figure \ref{ATF_3D})
reaches maximum at $\tau \sim 2.3$, while the average vertical component
of the dissipation rate (panel (b) in figure \ref{Diss_3D})
reaches maximum around $\tau \sim 2.8$.  Right after $\tau=2.8$,
we observe generation of vorticity inside the mixing zone, as shown
in panel (b) in figure \ref{HVty2V3}. Moreover, the horizontal component
of the advective fluxes (panel (a) in figure \ref{ATF_3D}) starts to
develop structures around $\tau=2.8$ as well. 

\begin{figure}
%\vspace{15pc}
\begin{center}
\epsfysize=2.0in
\epsfbox{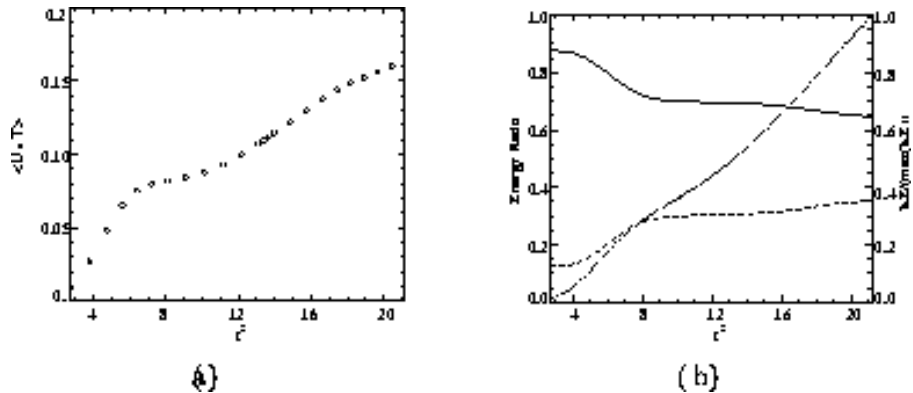}     
\end{center}
%\begin{center}
%\begin{tabular}{cc}
%{\psfig{file=cEntraintment_3D01_a01.eps,width=2.8in}} &
%{\psfig{file=cEnergyRatio_3D01_a01.eps,width=2.8in}} \\
%(a) & (b) \\
%(a) \\
%{\psfig{file=3DEnergyRatio_2D01_a.eps,width=2.5in}} &
%{\psfig{file=cEnergyRatio_3D01_a01.eps,width=3.5in}} \\
%(c) & (d) \\
%(b) \\
%\end{tabular}
%\end{center}
\caption{Panel (a): Volume-averaged advective fluxes as a function of time.
	 Panel (b): Partition of kinetic energy as functions of time.
         The solid line in panel (b) is the partition of
         the kinetic energy in
         the vertical direction and the dashed line is along
         the horizontal directions.  The dash-dotted line is
         the normalized kinetic energy as a function of time.}
\label{EnergyRatio}
\end{figure}

We can now summarize the
evolution of the 3-D RT instability as follows:  As the interfacial
random perturbations initiates some circulation (small vortices), the
seeded vortices increase in strength and begin free falling, accelerated
by the density contrast across the interface (as we discuss further
below). This is the free-falling phase, where the 
mixing zone is expanding due to a constant acceleration at the
boundaries, and inside the mixing zone,
the interaction between vortices is not strong enough to
alter the free falling dynamics. 
At $\tau^2 \sim 8$, the dissipation rate begins to
increase at a lower rate as a result of increasing mixing inside the
mixing zone. The development of structures is most vigorous as the
horizontal flow is enhanced and the viscous dissipation rate plateaus at
$\tau^2\sim 10$. Hence we mark $\tau^2=10$ as the beginning of the mixing
phase. From figure \ref{Diss_ratio}, we also observe that the end of
the mixing phase can be identified at $\tau^2\sim 16$, when the viscous
dissipation rate resumes positive growth. 
The existence of this mixing phase also
manifests itself in the potential energy release rate and the kinetic
energy partition (figure \ref{EnergyRatio}). 
The potential energy
release rate is defined as the volume-averaged, vertical advective flux
as follows:
\begin{equation}
\label{PEnergy_eq}
\int_{\Omega} w T d^3x =\partial_t\int_{\Omega} zT d^3x+
\kappa T|^{z=1}_{z=-1}.
\end{equation}
(We have utilized no-flux boundary conditions in deriving the above
equation.) The last term on the right hand side is negligible for small
$\kappa$, as the temperature difference is always of order one in our
simulation. In panel (a) of figure \ref{EnergyRatio}, we observe just
before the beginning of the mixing phase, the potential energy release
rate reaches a plateau.  In panel (b), the partitions of kinetic energy
remain more or less constant throughout the mixing phase, after which
they resume their original direction of varion with time.  Thus we
conclude that this mixing phase ($10\le \tau^2 \le 16$) is a regime where
horizontal components amplify and the lateral transport 
is also enhanced in the flow.
After this stage, the boundary of the mixing zone is not
pushed outward by a collection of plumes of the smaller sizes as in the
free-falling phase, but by a small number of big plumes 
that manage to survive the
mixing phase and expand in size as they propagate outwards.
It is therefore important for us to understand the evolution of plumes
and their propagation in various kinds of ambient background flows.

We also note that, based on our discussion, we expect to see
departures from simple $\tau^2$ scaling only when the dynamics is
dominated by a small number of (larger) plumes.  As a consequence,
the aspect ratio of the experimental domain can be decisive in 
determining whether such departures are actually observed: as the
height-to-width ratio decreases, we expect to see departures from
$\tau^2$ scaling at later and later times, so that plumes may begin
to interact with the box walls before such departures can be
observed.  Thus we conclude that the
dynamics and evolutions of single fingers (plumes) are important
to the broadening of the mixing zone in both
2-D and 3-D RT instability.
%
%Plumes in a nonstratified background fluid can 
%be regarded as a point source (sink)
%of buoyancy. 
%In the following sections, we focus on the dynamics and evolutions of plumes.
%Furthermore, if the interaction between plumes is weak, 
%each plume
%can be treated as an individual point source in a mean background flow.
%
In the following sections, we first discuss RT models based on the
point sources and interactions between each source.  We will introduce
the buoyancy-drag plume model in the context of the existent
point source models, and uncover the connections between
these models based on the physical grounds.

\section{Models for RT instability}
Early analytic work on modeling the mixing zone broadening goes back to
\cite{Fermi51} (see also \cite{Sharp84}); more recent models can be found
in \cite{Alon93} and \cite{Alon94}. From the numerical simulations of the
random-perturbation cases, we observe that plumes near the edge of the
mixing zone are  responsible for the broadening.  
We also observe that for the
parameters we adopt for our simulations, the interior flow of the mixing
zone is not homogeneous and is not isotropic. Instead, the flow pattern
inside the mixing zone has a preferred direction associated with the
plume structure inside the  mixing zone and cannot be described
as fully-developed turbulence.  
%We also observe how the
%propagation of the mixing zone transitions as the horizontal movement
%increases in amplitude. 
Motivated by these observations, we focus on RT
models that do not require the existence of turbulence to achieve 
the free-fall scaling.
We first briefly summarize two such models for immiscible RT
instability:  the first is for strongly stratified RT instability $(A=1)$;
the second is for weakly stratified RT instability
$(A\rightarrow 0)$.  The similarities between these two models (as will be
shown) lead one to conjecture that there may  exist a point-source
model that incorporates both limits.   Indeed, the long-existing
buoyancy-drag model, often used in modeling thermals or plumes,  seems to
capture the essential physical features in both models; and we will argue
that this model provides an adequate framework for capturing the 
essential physics of nonlinear RT instability.
We will discuss this third model at the end
of this section.

\subsection{Two models: Bubble competition model ($A=1$) \\
		    and Point source model ($A\rightarrow 0,\; Ag$ fixed).}

Zufiria's bubble competition model (\cite{Zufiria_87}) assumes a 2-D
potential flow model with an Atwood number of $1$.
In essence, this model is an extension of
Wheeler's model  (\cite{Sharp84}) to the 
strongly stratified limit.  The model consists
of variables that quantify the internal potential flow within each
bubble (point source inside the bubble)
and the distances of the source of the internal flow to the
bubble boundaries (radii of curvature of the bubble).
For a system of $20$ bubbles at the beginning, an estimated
range $0.04\le \alpha \le 0.07$ is obtained from this model.

A different model for 2-D inviscid, but weakly stratified RT instability,
%without the potential-flow assumption,
is given by \cite{Aref_Tryggvason_89}.
This model focuses on the Boussinesq limit, i.e., $A\rightarrow 0$ and
$Ag$ remains constant.  Following \cite{Tryggvason88}, the inviscid 2-D
Boussinesq RT instability is formulated in terms of vortex sheet strength
across the density interface.  Vortex sheet strength, defined as the
difference of the tangential velocity along the interface per unit length,
is driven by the density contrast across the interface.  In their model,
they are able to capture the complicated vortex pair evolution with
a simple evolutionary equation for the circulation per arc length.
They also successfully extend the single vortex pair model to
multi vortex pair systems, and the initial linear
growth rate from their model is in fair agreement
with the analytic results.

\subsection{Relating the models: The buoyancy-drag model}

As pointed out by \cite{Aref_Tryggvason_89}, their vortex model is
similar to the bubble-competition model: the point source in the bubble
model corresponds to the point vortices in the vortex model,
and the radii of curvature of the bubble correspond to the vertical
coordinates of the point vortices in the vortex model. The striking
similarities stem from the fact that bubbles, especially ones elongated in
the stratified direction, have flow patterns that amount to a vortex ring
near the elongated end and a point source inside the bubble feeding the
vortex ring. Therefore, the buoyancy-drag model used in modeling plumes
or thermals seems to be a natural candidate to connect the bubble picture,
where point sources are driven by a strongly stratified density contrast
across the interface, to the vortex picture, where the point sources are
driven by infinite gravitational acceleration (with $Ag$ constant).

The most simplified buoyancy-drag model (\cite{Lighthill_86,JWerne_94}) describes
the propagation velocity $(V)$ of a plume as follows:
\begin{equation}
\label{b_drag_01}
\frac{d (\rho_{in} V)}{dt} = \delta\rho\,\alpha g  - C_D \rho_{out} V^2,
\end{equation}
where $\alpha g$ is the effective acceleration due to the buoyancy
contained by the thermals to drive the motion, $\delta \rho\equiv
\rho_{out}-\rho_{in}$  is the density contrast between
the inside and outside of the
thermals, and $C_D$ is the drag  coefficient which depend on the
geometry of the thermals and the fluid viscosity.
There are more elaborate forms of equation
(\ref{b_drag_01}), such as different forms for $C_D$,
%and energetic 
%equations coupled to equation \ref{b_drag_01}, 
but the most essential feature is that if the
correction term to the constant acceleration is small when compared to the
acceleration, free-fall scaling is naturally obtained. 
%We further note
%that the second term on the right hand side of equation (\ref{b_drag_01})
%serves as a deceleration, as opposed to the first acceleration term.
Other variant forms of this simple model have been applied by
\cite{JKane98} and \cite{Dimonte_99} for a variety of parameters such as
Atwood number and Mach number, and the value of $\alpha$ is
claimed to be 
satisfactorily in agreement with results from  either experiments or
numerical simulations. 
In the next section, we perturb the RT interface 
with small, initial vortices: Such perturbations generate single
plumes expanding in size as they propagate away from the interface.
As will be shown, such plumes are well described by
equation (\ref{b_drag_01}). By comparing the propagation velocity
with the circulation of the vorticity, we are able to connect 
the point vortex model to the single plume model.  
We are also able to explain differences between 2-D and 3-D miscible
RT instability (either the random cases in \S 3 or the single plume
cases in \S 5) by conducting 
simple energetic diagnostics on the single plumes.

\section{The evolution of plumes}

\begin{figure}[h]
% \vspace{12.0pc}
\begin{center}
\epsfysize=1.8in
\epsfxsize=5.0in
\epsfbox{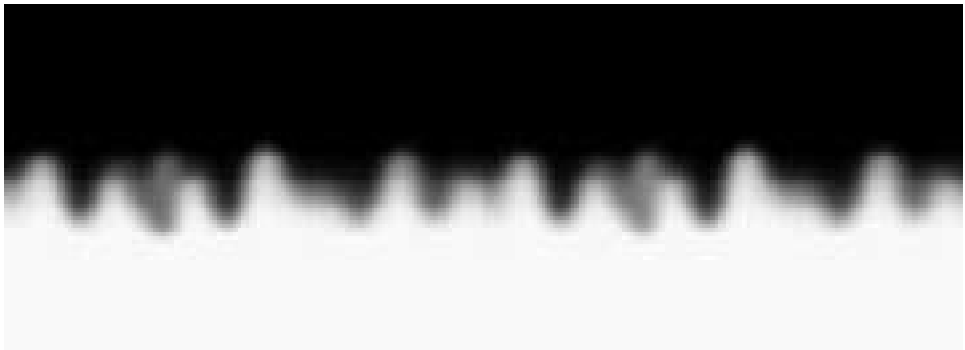}     
\end{center}
%\epsfxsize=5.0in
%\epsfysize=1.2in
%\epsfbox {demo02.ps}
\caption{Plume structures at the early development
of miscible Rayleigh-Taylor
instability.  Initially a layer of cold fluid (black) is
on top of a layer of warm fluid (white), and a random perturbation is
imposed at the interface.}
\label{Fig.demo01}
\end{figure}

The above discussion of the buoyancy-drag model for the nonlinear development
of the Rayleigh-Taylor instability, and the observation of Rayleigh-Taylor
``fingers" in the experiments (\S 3), 
strongly suggests an analogy between the
finger structures in the miscible version of the instability and the
evolution of isolated thermals or plumes. These latter structures have
been investigated extensively in the past (\cite{JSTurner}).
Experimentally, a bubble of hot fluid is injected near the bottom of a
container; the resulting buoyant plume expands as it rises and moves away
from the source. \cite{Libchaber93} have explored various situations,
which include injection of plumes into a turbulent flow. Experimental
observations show that as a localized source of buoyancy is injected
into the fluid, the resulting convection pattern consists of a cap which
forms at the outwards propagating front, and a stem which connects the cap
and the local source (\cite{Libchaber93}). In the case of the miscible RT
instability, the perturbation at the interface serves equivalently as an
injection of buoyancy sources (or sinks) into colder (hotter) fluids. The
early development of the perturbation then resembles a collection of
plumes going up and down from the interface (figure \ref{Fig.demo01}). 

As shown in this figure, each rising and falling bubble has a well-defined
cap and a stem connecting the cap to the root at the interface, consistent
with expectations for the behavior of plumes.

We have numerically simulated plumes which are initiated by perturbing a
Rayleigh-Taylor interface with a vortex pair in 2-D or a single vortex ring
in 3-D. Our interest is primarily in plume propagation, the temporal
development of the detailed geometry of the plumes, the dynamics and
energetics of the plumes, and the dimensionality of the plumes.
We remark in passing that 2-D plumes are referred to as planar vortex
pairs in three dimensions, i.e., three-dimensional plumes in which all
motions are restricted to lie in parallel, vertically-oriented planes.

In the following, we first establish a connection between miscible RT vortex
pairs and the immiscible version (\cite{Tryggvason88}) by comparing the
dynamics and the detailed vortex structures. We then present
results of two-dimensional simulations (with parameters listed in Table
2), and then go on to compare these results with the corresponding results
for three-dimensional calculations. The parameters for our single plume
simulations are listed in Table 2. The 2-D plumes are planar, while the
3-D plumes have cylindrical geometry. The initial perturbation imposed at
the interface is a Gaussian wave packet with amplitudes listed in Table 2.

\begin{table}
\label{Table1}
 \begin{center}
  \begin{tabular}{ccccccc}
Run number&$\beta g$&
$\nu\times 10^{3}$ & $\kappa\times 10^{3}$ & $d$ & $a$ & aspect ratio
\\[3pt]
    1 & 2.5    & 0.5 & 0.5 & 0.1 & 0.14 & $0.2:1$      \\
    2 & 2.5    & 1 & 1     & 0.1 & 0.14 & $0.2:1$      \\
    3 & 2.5    & 1 &  0.5  & 0.1 & 0.14 & $0.2:1$      \\
    4 & 2.5    & 0.5 & 0.5 & 0.1 & 0.11 & $0.2:1$      \\
    5 & 2.5    & 1 & 1     & 0.1 & 0.11 & $0.2:1$      \\
%    6 & 1.0   & 1 & 5     & 0.1 & 0.14 & $0.2:1$      \\
  \end{tabular}
\caption{Parameters for single plume simulations;
$d$ is the interface thickness, and $a$ is the amplitude of the
perturbation. }
\end{center}
\end{table}

\subsection{2-D single plumes}

\subsubsection{Connection to weakly stratified, immiscible R-T instability}

\cite{Tryggvason88}
utilized the vortex sheet strength to describe  the evolution of single
mode perturbations in 2D immiscible RT instability.  In the case of
miscible R-T instability, one can construct a family 
of isothermal contours due
to the diffusivity; we can monitor the evolution of single
plumes by evaluating various quantities (such as vorticity, thermal flux)
on the isothermal (or, in Boussinesq fluids, isodensity) contours.
%\clearpage
\begin{figure}[htbp]
%\vspace{25pc}
\begin{center}
\epsfysize=3.0in
\epsfbox{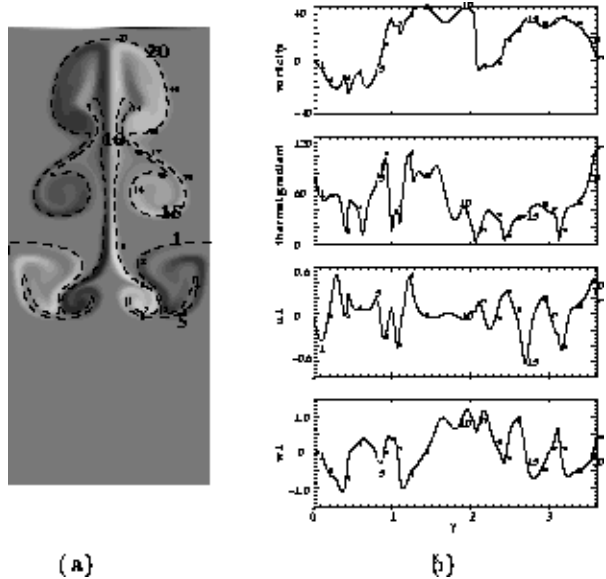}     
\end{center}
%\begin{center}
%\begin{tabular}{c c}
%{\psfig{file=2Dsingle_plume_vort02.eps,width=2.5in,height=4.0in}} &
%{\psfig{file=2Dsingle_plume_lq02.eps,width=2.5in,height=4.0in}} \\
%(a) & (b) \\
%\end{tabular}
%\end{center}
\caption{2D vorticity contour plot (panel (a)) and the value of
various physical quantities
along the $T=0.75$ isothermal curve (panel (b)).
In panel (b), vorticity (first plot), the projection of the thermal
gradient on the isocurve (second plot), the horizontal
heat flux (third plot) and the vertical heat flux
(fourth plot) are evaluated along the isocurve. For
illustration, we place numbers along the isocurve to display the
important spatial features related to the flow pattern.}
\label{AP_2Dsingle_plume}
\end{figure}
As long as the structure of the isothermal contours remains relatively
simple (as shown in panel (a) of figure \ref{AP_2Dsingle_plume}),
we can gain some understanding of the structure of the buoyant plumes by
considering the evolution of various physical quantities along the
isothermal surfaces. Panel (a) shows the spatial structure of the planar
vortex pair for the first set of parameters in Table 2. 
The roll-up structure is reminiscent of those observed in the weakly
stratified, immiscible R-T instability (\cite{Aref_Tryggvason_89}). To
make complete comparison, we first introduce the circulation ($\Gamma$)
along some closed curve:  The circulation $\Gamma$, defined as the line
integral of the flow along a closed curve,
\begin{equation}
\label{def_circ}
\Gamma \equiv \int_{\bf C} \vec{v}\cdot d\vec{l},
\end{equation}
is related to the evolution of vortex pair as follows:
\begin{equation}
\label{circ_vort}
\frac{d \Gamma}{d t} = \int\int_{\bf A}
\nabla\frac{1}{\rho}\times\nabla p\cdot d{\bf A},
\end{equation}
where ${\bf A}$ is the surface enclosed by the curve ${\bf C}$.
In the case of weakly stratified Rayleigh-Taylor instability
(\cite{Aref_Tryggvason_89}),
the above relation is modified by replacing the
right-hand-side integral with the potential
energy density due to the interfacial density contrast.
The propagation velocity $(V)$ of the vortex pair
in a nonstratified fluid 
is then related to the circulation $\Gamma$ as
\begin{equation}
\label{v_vort}
V\;\;=\;\; \Gamma/\eta,
\end{equation}
where $\eta $ is basically the correlation length for vortices,
and is a function of geometry and dimensionality:
one expects $\eta $ for 2-D vortex pairs
to be larger than for 3-D vortex rings.
Equation (\ref{v_vort}) is valid even in the case of weakly stratified
R-T instability, where vortices, originally initiated at the interface
due to the density perturbation, propagate in a uniform ambient background.
%As the vortex pair moves away from the initial interface,
%the circulation evolves according to equation \ref{circ_vort}.

With equation (\ref{v_vort}), we can connect the bubble model
to the point vortex model via the buoyancy-drag plume model.
To fix ideas, we first outline an  evolutionary scenario for the weakly
stratified, miscible R-T instability: As we disturb the interface with
some temperature perturbation, vortices are generated as a result of
lateral thermal gradient.  Thus some initial circulation is seeded inside
the vortices and a propagation velocity is acquired according to equation
(\ref{v_vort}). The circulation then evolves according to equation
(\ref{circ_vort}), and the propagation velocity is approximately given
by equation (\ref{v_vort}) at each time step for an updated circulation
(\cite{Hill_75}). We also observe that the acceleration experienced 
by the vortex pair
is  proportional to the increase/decrease rate of change of 
circulation $\Gamma$. Combining equations (\ref{circ_vort}) and
(\ref{v_vort}), we obtain the equation for the acceleration for the
planar vortex pair:
\begin{equation}
\label{vort_acc_s}
\frac{d V}{d t} \;\;=\;\; \frac{d (\Gamma/\eta )}{dt} 
                \;\;=\;\; g f(\delta\rho, \hat{\bf s}),
\end{equation}
where $g$ in equation (\ref{vort_acc_s}) is the gravitational acceleration,
and $f$ is a function of density contrast ($\delta \rho$)
and geometry of the interface
(unit tangent vector along the interface $\hat{\bf s}$).

For point vortices (\cite{Aref_Tryggvason_89}),
$f = {\hat g}\cdot \hat{\bf s}$ where ${\hat g}$ is the unit vector
along the gravitational direciton and $\hat{\bf s}$ is the unit tangent
vector along the interface.
In our case, we expect $f$ to be a constant
(effective acceleration) plus some correction term due to viscous drag.
This expectation rises from two facts: The first is
the observation that
the plume consists of a vortex pair inside the cap and
a flow sustaining the vortex structure (the neck connecting
the cap to the stem).  Therefore, we expect the dynamics of the
vortex pair to be similar to the plume dynamics.
%where the plume is not sustained by a constant source.
The second fact is that the
propagation of buoyant plumes
(not sustained by a constant source)
in a homogeneous fluid is described
by the buoyancy-drag model (equation \ref{b_drag_01}), where
the constant density difference across the plume boundary provides buoyancy
acceleration,
and the viscous drag experienced by the plume causes a deceleration.
To be more specific, we expect $f$ to take the following form:
\begin{equation}
f = \alpha + \gamma(\nu,\delta \rho, V),
\end{equation}
where the ``constant" $\alpha$ may depend on the dimensionality and the
initial perturbation, and the correction terms are included in the
function $\gamma$ (a function of viscosity $\nu$, the propagation velocity
$V$ and the density difference $\delta \rho$).
If $\gamma$ is chosen to be proportional to $-V^2$ as in \cite{JWerne_94},
the propagation velocity will reach a constant terminal
velocity ($V=V_f$ is a constant) after a free falling period ($V\propto
t$ and $h\propto t^2$). 
%Of course, in the free falling regime
%the effective acceleration constant $\alpha$ is of most interests.

To fully establish the correspondence between the immiscible point
vortex model and the miscible plume model,  we calculate the circulation
$\Gamma$ along isothermal contours (not necessarily closed) in our 2-D
simulations. 
\begin{figure}
%\begin{center}
%\begin{tabular}{c c}
%{\psfig{file=c2d_single_plume_circulation.eps,width=2.5in}} &
%{\psfig{file=cPlume_model_fit01.eps,width=2.5in}} \\
%(a) & (b) \\
%\end{tabular}
%\end{center}
%\vspace{15pc}
\begin{center}
\epsfysize=2.0in
\epsfbox{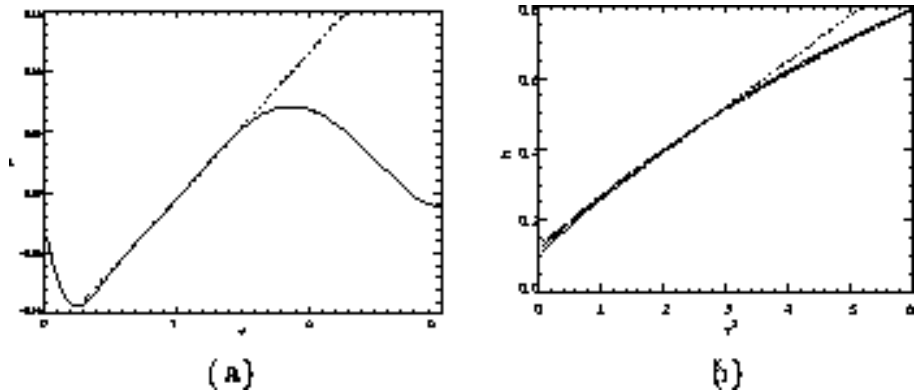}     
\end{center}
\caption{Circulation $\Gamma$ (along the $T=0.7$ isothermal contour)
	 and the propagation of a 2-D single plume. In panel (a),
         the solid line is from the simulation, and the dashed line
         is the $\tau^2$ fit. In panel (b), the
         solid line is from the simulation, the dashed line is the
         $\tau^2$ fit, and the dash-dotted line is the fit to
        the single plume model.}
\label{circu_depth}
\end{figure}
Figure \ref{circu_depth} (a) shows the temporal evolution
of $\Gamma$ along $T=0.7$, and during $0.7<\tau <2.1$, we find $\Gamma$ to
be proportional to $\tau^2$. Comparing with figure \ref{circu_depth} (b),
we note that this duration coincides with the free falling phase when $h$
scales to $\tau^2$.   This is similar to what is observed in
\cite{Aref_Tryggvason_89}: the propagation of the 2-D vortex pair is
free-falling at the beginning, and it decelerates due to
the decrease of available potential energy per unit arclength as the rollup
continues 
(and also due to diffusion and viscous dissipation in our case) - 
thus the decrease in circulation inside the vortex pair.  We also
note that, the constant initial acceleration rate of $\Gamma$
even in the miscible case indicates that the initial
circulation is due to the buoyancy-driven flow around the cap
head. As the circulation increases, drag becomes important
and the circulation saturates to a maximum. The circulation then
decreases at late times due to the strong enhancement of the horizontal
motion, which is typical of 2-D flow and less so in 3-D flow
(we will discuss this difference later). We should point out here that,
despite the similarities already presented between point-vortex model and
plume model, there exists an essential difference: the plume has a stem
structure associated with the circulation around the neck,
which closely resembles the flow in the vicinity of RT ``fingers".
Our future research problem is to understand how the presence of
multiple plumes affects the circulation near the necks; this will
be an important future step in connecting these models to the 
RT finger evolution in more vigorous flows (higher Reynolds numbers).
%This distinguishing factor is important concerning the evolution
%of multiple plumes. The propagation of each plume is
%influenced by the flow near its stem. For example,
%in the case of multi-plume system (an analogy to the random
%case in \S 3), the deceleration each plume experiences can be amplified
%due to the enhanced horizontal shear motion around the stems
%(as shown in \S 3). The detailed coupling mechanism can be
%model-dependent, and remains to be an interesting future direction.

We are now ready to compare the vortex structure between the
miscible R-T instability and the immiscible version. 
From panel (b) of figure \ref{AP_2Dsingle_plume},
we obtain the following observations:

\noindent
{\bf (1)}
The vorticity vanishes at the tip of the cap head. This implies that the
flow near the cap is essentially irrotational, and explains why the potential
flow assumption adopted in the derivation of the cap shape by
\cite{Lighthill_86}
(see also \cite{Libchaber93}) works so well in describing the head shape.
This is also consistent with results for the weakly stratified, inviscid
R-T instability (\cite{Aref_Tryggvason_89,Tryggvason88}).

\noindent
{\bf (2)}
The vorticity reaches extrema near the neck of the plumes, i.e., near
the part of the plume which connects the stem and the cap head; this is
the location where circulation is most vigorous.  This is also consistent
with the fact that circulation reaches extremum at the inflection points
along the interface (\cite{Aref_Tryggvason_89}).

\noindent
{\bf (3)}
The amplitude of the thermal gradient reaches extrema either near the
cap heads or near the neck of the plumes. This is consistent with the
previous observation, because the larger the thermal gradient is, the
larger the rate of circulation is, and thus the vorticity reaches
extrema at these points.

\subsubsection{Dynamics and energetics of 2-D plumes}
\label{DnE_2Dplumes}
In this subsection, we study the details of the single plume 
structures to uncover the physical ingredients essential to 
model the RT fingers.
As in the point vortex model, we find the evolution of arclength
of isothermal contour to be indicative of the detailed evolution
of the plume structures.  Here 
we first focus on the evolution of the isolevels associated with single
plumes. Secondly we explore the dependence of the single plume
evolution on the various parameters such as the amplitude of the
perturbation, the Reynolds number and the Prandtl number.

\begin{figure}[ht]
%\vspace{15pc}
\begin{center}
\epsfysize=2.0in
\epsfbox{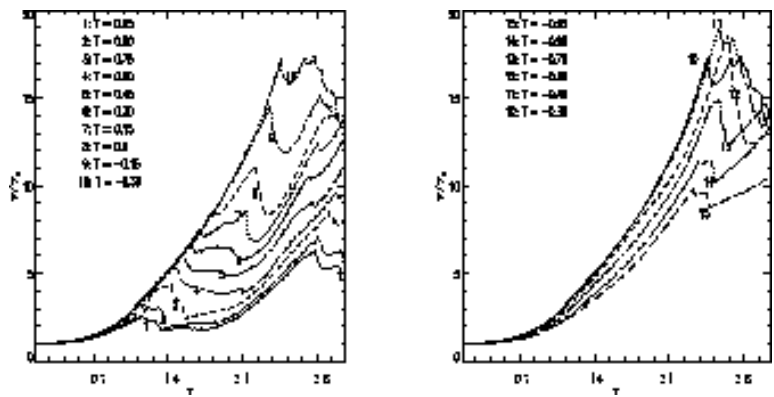}     
\end{center}
%\begin{center}
%\begin{tabular}{c c}
%{\psfig{file=c2Dsingleplume_arclength01.eps,width=2.5in,height=2.0in}}
% &
%{\psfig{file=c2Dsingleplume_arclength02.eps,width=2.5in,height=2.0in}}
%  \\
%\end{tabular}
%\end{center}

\caption{Amplification of arc length as a function of time for $15$
different isolevels (2D).}
\label{AL_2Dsingle_plume}
\end{figure}
Figure \ref{AL_2Dsingle_plume} is an example of how
the arc length evolves as a function of time for different levels from
the 2-D simulation (run number 4 in Table 2).  
These are the time evolution of arc
length for $15$ different isothermal contours. These various evolutions
in the isothermal contours can shed  light on the detailed, local
structure of the fluid. First we observe that, among all the $15$ curves,
curve $11$ ($T=-0.4$) has the longest curve length. Curves $1-6$ in
figure \ref{AL_2Dsingle_plume} correspond to isolevels near the boundary
between the plume and the ambient fluids. Before
$\tau\sim 1.0$, the flow inside the plume pushes these levels upward as
the plume rises.  From $\tau\sim 1.0$ on, these contour levels begin to
catch up with the ascending   boundary of the plume to the heavy fluid,
and the level sets break up into several smaller connected loops. After
this event (at $\tau\sim 1.4$),  the plume has established its own shape
and the cap propagates in a shape-preserving fashion.
At time $1.7 < \tau < 2.1$, small vortex pairs near the original interface
begin to roll up to  bigger vortex pairs and thus the increase in arc
length. Curves $7-11$ represent transition contours inside the plume.
The sudden drop at $\tau\sim 2.1$ correspond to the shedding of  smaller
vortex pairs.  Curves $12-15$ represent the evolution of the levels near
the boundary of the ascending plumes, and peaks in each curve represent
break-offs of isolevel curves. 

\begin{figure}[ht]
%\vspace{15pc}
\begin{center}
\epsfysize=3.5in
\epsfxsize=3.0in
\epsfbox{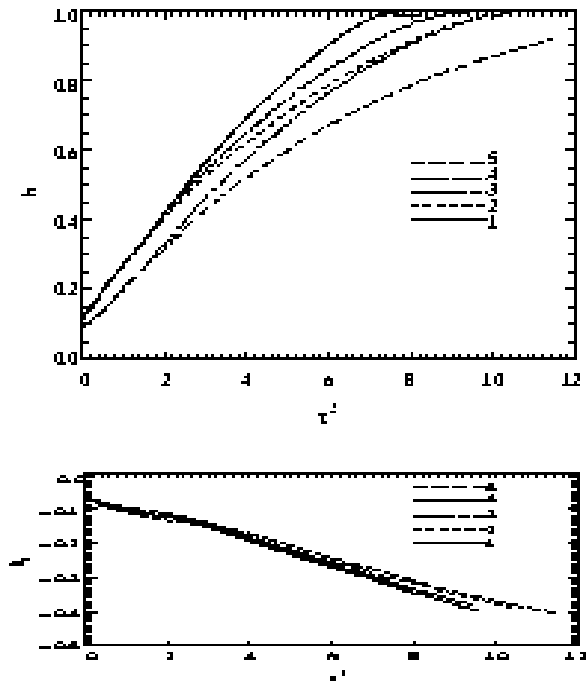}     
\end{center}
%\begin{center}
%\begin{tabular}{c}
%{\psfig{file=c2Dsingle_plumes_h01_a01.eps,height=2.4in,width=3.0in}} \\
%{\psfig{file=c2Dsingle_plumes_h02_a01.eps,height=1.2in,width=3.0in}} \\
%\end{tabular}
%\end{center}
 \caption{Penetration depth versus $\tau^2$ for the 2D single plumes.
  The top panel is for the ascending boundary and the bottom
  panel is for the descending boundary. The numbers are the
  run numbers listed in Table 2.}
 \label{h_2Dsingle_plume}
\end{figure}

We now focus on how the penetration depth
$h$ grows in time, and how this temporal evolution depends on various
control parameters in the simulations. Figure
\ref{h_2Dsingle_plume} shows curves of $h$ versus $\tau$ for our 2-D
simulations. We first note that at early time, the ascending $h$
(top panel of figure \ref{h_2Dsingle_plume}) does scale with $\tau^2$ from
$\tau^2\sim 0.2$ to $\tau^2\sim 3.0$.   
After $\tau^2=3.0$, the plumes
have reached a terminal velocity  and therefore travel at a constant
speed, which is often observed in plumes without sustaining sources. We
further note that the higher the Reynolds number,  the faster the plume
penetrates (cf. curve 1 and curve 2). We also observe faster penetration
as we increase the diffusivity ratio $(\nu/\kappa)$  (cf. curve 2 and
curve 3), which  is a direct result of weaker stabilization for smaller
diffusivity. We have also varied the initial perturbation amplitude (cf.
curves 1 and 4, and curves 2 and 5): For perturbation of amplitude below
$8\%$ of the total box height, we find that  the larger the initial
perturbation, the faster the plume penetrate. The penetration depth
depends linearly on the initial amplitude until $\tau \sim 1.4$, where
nonlinearity may have set in to amplify the dependence on the initial
state (cf.\ curve $2$ and curve $5$).  We also note that the penetration
of the heavy fluid into the light fluid remains more or less  similar as
we vary these parameters. The strong dependence of the ascending
structure on the parameters is not reflected on the dynamics of the
descending structures. This demonstrates that, with this form of initial
perturbation, the ascending structures have different dynamics than the
descending  structures despite the low Atwood number adopted under the
Boussinesq approximation.  This is certainly not the case if we perturb
the interface with symmetric disturbances. 
%and therefore, serves well to
%demonstrate the effect of initial perturbation on the miscible RT
%instability.

\subsection{Comparison between 2-D and 3-D plumes}

\begin{figure}[ht]
%\vspace{15pc}
\begin{center}
\epsfysize=3.0in
\epsfxsize=4.0in
\epsfbox{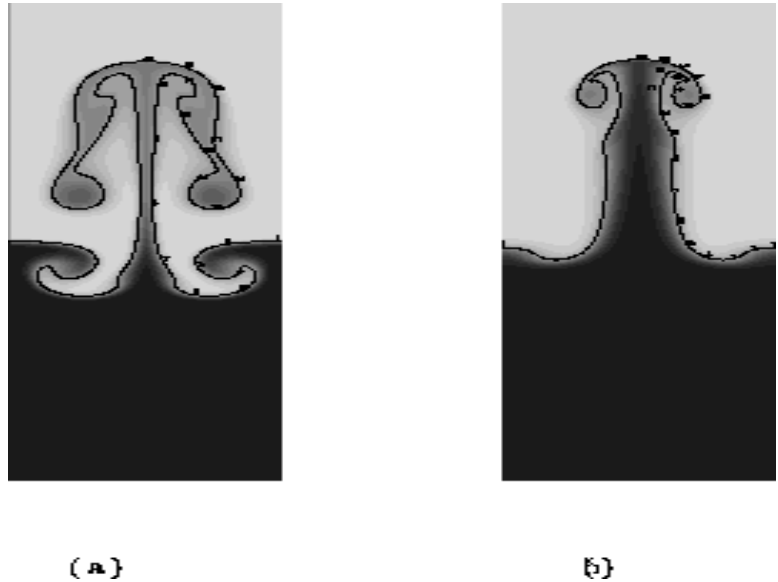}     
\end{center}
% \begin{center}
%   \begin{tabular}{c c}
%    {\psfig{file=2dsingleplume_isocurve01.eps,height=4.5in,width=2.25in}} &
%    {\psfig{file=3dsingleplume_isocurve01.eps,height=4.5in,width=2.25in}} \\
%    (a) & (b) \\
%   \end{tabular}
% \end{center}
\caption{Comparison of 2D (panel (a))
         and 3D plume (panel (b)) structures: geometric shape. The
         solid line is the $T=0.7$ isothermal contour.}
\label{2d_vs_3d_spstructure}
\end{figure}
\begin{figure}
%\vspace{15pc}
\begin{center}
\epsfysize=2.0in
\epsfbox{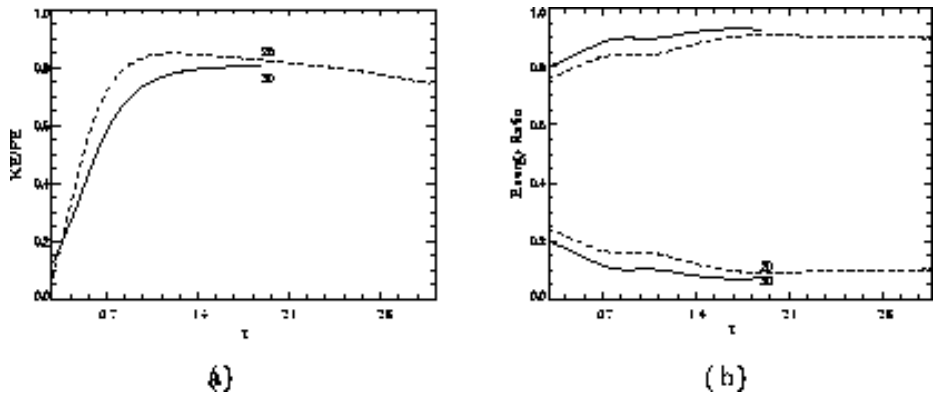}     
\end{center}
% \begin{center}
%   \begin{tabular}{c c}
%     {\psfig{file=c2dVS3d_plumes_energetic02.eps,width=3.0in}} &
%     {\psfig{file=c2dVS3d_plumes_energetic04.eps,width=3.0in}} \\
%     (a) & (b) \\
%   \end{tabular}
% \end{center}
\caption{Panel (a): Ratio of total kinetic energy to total potential
	 energy available in the system as a function of $\tau$.
	 Panel (b):  Vertical partition of the total kinetic energy and
	 horizontal partition of the kinetic energy as functions of $\tau$.
	 In both panels,
	 we plot 2D (dashed lines) together with 3D (solid lines).}
\label{2d_vs_3d_spenergetics}
\end{figure}

             
In this section we present results from simulations of
both 2-D and 3-D single plumes for the first set of
parameters in Table 2.      
The differences between 2-D and 3-D
plumes are manifested in various
aspects.  In figure \ref{2d_vs_3d_spstructure}
we display the plume structures (at
the same penetration depth) in both
2-D and 3-D. 
%\begin{figure}[ht]
% \begin{center}
%   \begin{tabular}{c c}
%    {\psfig{file=2dsingleplume_isocurve01.eps,height=4.5in,width=2.25in}} &
%    {\psfig{file=3dsingleplume_isocurve01.eps,height=4.5in,width=2.25in}} \\
%    (a) & (b) \\
%   \end{tabular}
% \end{center}
%\caption{Comparison of 2D (panel (a))
% and 3D plume (panel (b)) structures: geometric shape. The
% solid line is the $T=0.7$ isothermal contour.}
%\label{2d_vs_3d_spstructure}
%\end{figure}                  
We also compute various quantities along the isothermal contours,
and basically the
same spatial structures are found in the 3-D plumes:
the vorticity goes to zero at the cap head,
the maximum circulation
occurs around the neck, and positions of
where temperature gradient have extrema are
also points where the vorticity have extrema.
The only discernible difference
between 2-D and 3-D is the
magnitude: the 3-D thermal gradient has
higher maximum at the cap head,
and the advective heat transport
(both vertical and horizontal components) in 3-D
is much smaller compared to the 2-D heat transport.
The latter (difference in the efficiency of heat
transport) explains the thinning of
the 2-D stem as opposed to a constant
width of the 3-D stem observed in our simulations.
From our simulations, we also
conclude that, for the same parameters,
the cylindrically symmetric plumes (our 3-D plumes) travel
much faster than the planner plumes (our 2-D plumes).
\begin{figure}
%\vspace{15pc}
\begin{center}
\epsfysize=2.0in
\epsfbox{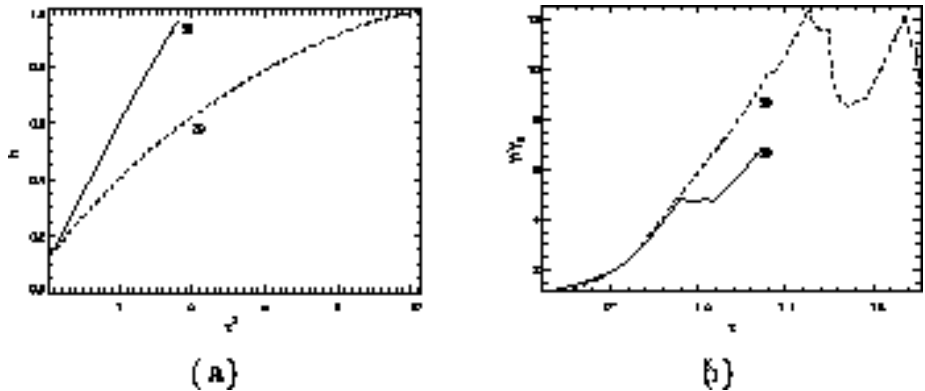}     
\end{center}
% \begin{center}
%   \begin{tabular}{c c}
%     {\psfig{file=c2dVS3d_plumes_height01.eps,width=2.5in}} &
%     {\psfig{file=c2dVS3d_plumes_arclength01.eps,width=2.5in}} \\
%     (a) & (b) \\
%   \end{tabular}
% \end{center}

\caption{Penetration depth (panel (a)) and
	 the arclength of $T=0$ isothermal curve (panel (b)):
         2D (dashed lines) versus 3D (solid lines).}
\label{2d_vs_3d_spheight}
\end{figure}
As seen in figure \ref{2d_vs_3d_spheight},
the penetration depth in 3-D increases faster with
time than in
the 2-D case, and eventually reaches
almost 2 times the 2-D penetration
depth at the end
our our simulations. In order to
understand this difference, we have
calculated the 
temporal evolution of
the ratio of total kinetic energy to the
potential energy, and the partition of 
kinetic energy
(which we break up into its two
components, namely the total kinetic
energy
associated with vertical motions and
with horizontal motions).
Thus, in figure
\ref{2d_vs_3d_spenergetics} (a) we plot the
energetic ratio
versus $\tau$,
and show the evolution of the kinetic energy partition 
as a function of time in
figure \ref{2d_vs_3d_spenergetics} (b).
First, we point out that the rate of
conversion of potential energy to
kinetic energy is
roughly the same in 2 and 3-D.
Secondly, we note that
partitioning of the kinetic energy
into horizontal and vertical components
is different
in 2 and 3-D (with a larger fraction
of the kinetic energy going to
horizontal motions
in the 2-D case).
Based on these facts, we can now
attribute the increased effectiveness of plume
penetration in three dimensions to
the fact that while more potential energy is
released per unit time in
2-D than in 3-D, the partitioning of the
resulting kinetic energy into its components drastically
favors the directed vertical component in 3-D, but not
in 2-D. 
into the vertical
component of the flow kinetic energy, so
that the 3-D
plumes go faster and penetrate deeper
in the same amount of time than their 2-D
counterparts.
\subsection{Application of the single-plume model and future direction}
We apply the plume model to our single plume simulations and find
``good" agreement as we fit the penetration depth $h$ to the
model (in the sense that the absolute error is less than $1\%$ over the
whole evolution).
A typical example is shown in figure \ref{circu_depth}(b),
where the dashed line is the
least square fit for the free-falling law, and the dash-dotted
line is the least square fit for the plume model.
We clearly see that
the plume model captures the whole evolution, and the deviation
is small compared to the $\tau^2$ fit.
\begin{figure}[ht]
%\vspace{15pc}
\begin{center}
\epsfysize=2.0in
\epsfbox{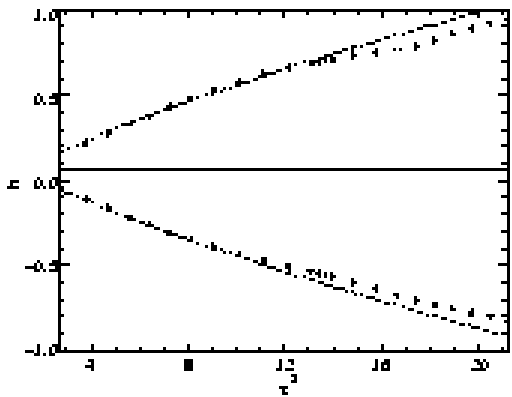}     
\end{center}
%\begin{center}
%\begin{tabular}{c}
%{\psfig{file=cs128erf_3d_h01_b01.eps,width=2.5in}}
%\end{tabular}
%\end{center}
\caption{The 3D random case revisited: the diamonds are from the
	 simulations and the dashed lines are the fits of the
         analytic plume model to the data during the first free-falling
         phase.}
\label{Penetration3_plumefit}
\end{figure}                
When we apply the plume model to the random case
\ref{Penetration3_plumefit},
the existence of
three evolutionary phases prohibits a
physically meaningful fit for the whole
duration of the evolution, and a reasonably good fit
is found for early evolution:
$2.6\le \tau^2\le 13.4$, where the effective acceleration from
this fit is $\alpha\sim 0.03$.
%\begin{figure}[ht]
%%\vspace{15pc}
%\begin{center}
%\begin{tabular}{c}
%{\psfig{file=cs128erf_3d_h01_b01.eps,width=2.5in}}
%\end{tabular}
%\end{center}
%\caption{The 3D random case revisited: the diamonds are from the
%	 simulations and the dashed lines are the fits of the
%         analytic plume model to the data during the first free-falling
%         phase.}
%\label{Penetration3_plumefit}
%\end{figure}
As shown in figure \ref{Penetration3_plumefit}, the evolution
deviates from the fit at $\tau^2\sim 11$, right
after the beginning of the mixing phase as discussed in
\S 3.  This is an indication that the interaction of plumes with the
background flow is substantial enough to arrest the free falling
propagation of the mixing zone.
To incorporate this interaction
with the background flow into the single plume model, one
needs an equation for the energy evolution of the
plume, and also an additional momentum term in equation \ref{b_drag_01}
to describe the momentum transfer between the horizontal motion
and the vertical motion.
The interaction between plumes,
such as merging, also has important contribution to the
evolution of plumes as they expand in size and getting closer
to each other.
This is a challenging future direction and
is currently under investigation.

\section{Summary and Conclusions}

We have performed numerical simulations in both 2-D and 3-D with
relevant length scales arranged such that the thickness of the interface
between the two fluids remains the same throughout the simulations.
We have found dramatic differences between 2-D and 3-D flows with the aid
of average quantities.  

We first perturb the RT interface with ``random" perturbations and
investigate the evolution of the mixing zone in both 2 and 3-D
with the aid of averaged quantities.
Scaling of the mixing zone width with $\tau^2$ is found in
both 2 and 3-D, but only during selected intervals of time.  
Moreover, we find the 3-D mixing zone expands
two times faster than the 2-D mixing zone.  An important
finding in this respect is that we can identify three phases
of evolution for the 3-D mixing zone:
the first is the free-falling phase right after
the linear growth period (after $6$ e-folding times, \S 3), the second
is the mixing phase where the free falling slows down and more mixing
is generated as a result of enhanced horizontal motions, as illustrated
in \S 3; the third is another free falling phase during which the broading
of the mixing zone resumes the $\tau^2$ scaling, when the big
plumes near the edge become decoupled from the mixing zone
and propagate on their own.  Whether all of these phases are observed
is likely to depend on the aspect ratio of the experimental domain.
In the
2-D single plume case, we explore the parameters to demonstrate the
effect of initial conditions on the broadening of the
mixing zone and thus the effective acceleration $\alpha$.
Through energetic analysis, we are able to attribute the difference
between 2 and 3-D flows to the fact that in 2-D, more energy
goes into highly correlated horizontal and vertical
motion (i.e., vortical motions) than into directed motions along
the gravitational direction (3-D).                   

We have discussed two analytic Rayleigh-Taylor models relevant to our
computations (the bubble-competition model (\cite{Zufiria_87}) and the
point-vortex model (\cite{Aref_Tryggvason_89}), which are effectively
two extreme limits of the  2-D incompressible RT instability from the
point-source point of view. The essential physical features found in both
models are also key constituents of the buoyant-drag plume model,
in which each single plume experiences an acceleration due to the density
contrast, and also a deceleration, which is due to a combination of viscous drag
and the horizontal shear flows in the ambient background.
We discuss how the plume model serves to
connect the two models on physical grounds, and point out that
the plume model actually contains more physical features, such as the
coupling of the plumes with the mixing zone via the interaction between a
background flow (inside the mixing zone) and flows in the stems of
the plumes (whose heads are pushing the envelope of the mixing zone).
Furthermore, we show that  for the miscible vortex pair,
both circulation $(\Gamma)$
and penetration depth $(h)$ scale with $\tau^2$
for the same duration of time. This illustrates the close connection
between the plumes and the point vortex pairs.  In addition, we also
strengthen the connection by comparing the evolution of RT interface in the
miscible case with that in the immiscible case. The extension of the
single plume model to multi-plume systems is of great interest and is
now under investigation.                                               

The success of the buoyant-drag model
is an indication of the fundamental physical ingredients of the RT
instability.  Each bubble can be considered as an internal point
source, and the curvature of the bubble evolves according to the
internal source strength and the 
relative position of the internal source to the bubble boundary.
The internal point source is fed by the flow from below
the bubble, namely the stem connecting the cap head of the bubble.
The coupling of the stem with the background flow, then, determines
the strength of the internal source.  All these features
are ingredients of the plume model, which is, strictly speaking,
an empirical model for plumes 
in both laminar and turbulent flows.
Therefore, we find the buoyant-drag
plume model to 
give an adequate description of the various physical processes
that we have observed.  The extension of this single plume model to incorporate
the coupling of plumes, merger events, and the background flow interaction
is challenging and will no doubt bring us more insight into the RT
instability.

\begin{acknowledgments}
\end{acknowledgments}
This work was supported as part of the validation program of the Center for
Astrophysical Thermonuclear Flashes, supported by the DOE/ASCI Alliances
program at the University of Chicago. We would like to acknowledge
extensive discussions of the Rayleigh-Taylor problem with N.J. Balmforth,
J. Biello,
A. Cook, G. Dimonte, B. Fryxell and L. Howard. The motivation and
stimulation were provided by the ``alpha group" project, led by G.
Dimonte.

%\appendix
%\section{}

\end{document}